\documentclass[superscriptaddress,twocolumn,showpacs,a4paper,
amssymb,amsmath,nobibnotes,aps,prd,
showkeys,
nofootinbib,notitlepage]{revtex4-1}
\usepackage{verbatim}
\usepackage[T1]{fontenc}
\usepackage[utf8]{inputenc}
\usepackage[american]{babel}
\usepackage{epsfig}
\usepackage{booktabs}
\usepackage{multirow}
\usepackage{dcolumn}
\usepackage{amsmath}
\usepackage{mathtools}
\usepackage{amsfonts}
\usepackage{amssymb}
\usepackage{ulem}
\usepackage{epstopdf}
\usepackage{bm}
\usepackage{siunitx}
\usepackage{braket}
\usepackage{enumitem}
\usepackage{soul}
\usepackage[table]{xcolor}
\usepackage{color}
\usepackage{transparent}
\usepackage{pifont}
\usepackage{enumitem}

\definecolor{navyblue}{rgb}{0.0, 0.0, 0.5}
\definecolor{royalblue}{rgb}{0.25, 0.41, 0.88}
\definecolor{cadmiumgreen}{rgb}{0.0, 0.42, 0.24}
\definecolor{blue-violet}{rgb}{0.54, 0.17, 0.89}
\definecolor{darkviolet}{rgb}{0.58, 0.0, 0.83}
\definecolor{orange(colorwheel)}{rgb}{1.0, 0.5, 0.0}

\usepackage{hyperref}
\hypersetup{
    colorlinks=true, 
    linkcolor=royalblue, 
    citecolor=magenta}

\usepackage{booktabs}
\usepackage{multirow}
\usepackage{dcolumn}
\usepackage{colortbl}

\begin{document}

\title{Dynamical Dark Energy Beyond Planck? Constraints from multiple CMB probes, DESI BAO and Type-Ia Supernovae}

\author{William Giar\`{e}}
\email{w.giare@sheffield.ac.uk}
\affiliation{School of Mathematical and Physical Sciences, University of Sheffield, Hounsfield Road, Sheffield S3 7RH, United Kingdom \looseness=-1}

\date{\today}

\begin{abstract}
\noindent 
The latest Baryon Acoustic Oscillation (BAO) measurements from the Dark Energy Spectroscopic Instrument (DESI) collaboration, when combined with Planck satellite Cosmic Microwave Background (CMB) data and Type Ia Supernovae, suggest a preference for Dynamical Dark Energy (DDE) at a significance level ranging from $2.8\sigma$ to $4.2\sigma$. In this work, I test whether, and to what extent, this preference is supported by CMB experiments other than Planck. I analyze the latest Atacama Cosmology Telescope (ACT) and South Pole Telescope (SPT) temperature, polarization, and lensing spectra at small scales, eventually combining them with Planck or WMAP 9-year observations at large angular scales. My analysis shows that ACT and WMAP data, when combined with DESI BAO and Pantheon-plus Supernovae, yield independent constraints with a precision comparable to Planck. Notably, in this case, the cosmological constant value is recovered within two standard deviations. A preference for DDE reappears when Pantheon-plus is replaced with distance moduli measurements from the Dark Energy Survey Supernova program (DESy5). However, it remains less pronounced compared to the Planck-based results. When considering SPT data, no clear preference for DDE is found in combinations involving Pantheon-plus Supernovae, and the preference is significantly weaker in combinations involving DESy5. Overall, CMB experiments other than Planck generally weaken the evidence for DDE. I argue that the subsets of Planck data that strengthen the shift toward DDE are the temperature and E-mode polarization anisotropy measurements at large angular scales $\ell \lesssim 30$.
\end{abstract}

\maketitle

\section{Introduction}
\label{sec:introduction}
The Dark Energy Spectroscopic Instrument (DESI) has recently released measurements of Baryon Acoustic Oscillations (BAO) based on its second year of observations (DR2), which include data from over 14 million extra-galactic objects, such as galaxies, quasars, and Lyman-$\alpha$ forest tracers~\cite{DESI:2024mwx,DESI:2024uvr,DESI:2024kob,DESI:2024lzq,DESI:2024aqx,DESI:2025zgx,DESI:2025zpo,DESI:2025qqy}. DESI BAO provide precise constraints on the transverse comoving distance, the Hubble rate, and their combination (all relative to the value of the sound horizon at the drag epoch) across seven redshift bins spanning a range $0.1 < z < 4.2$. 

One of the most important -- yet preliminary -- results emerging from these observations concerns the nature of Dark Energy (DE).\footnote{Other potential momentous implications concern neutrino cosmology. The upper limit on the total neutrino mass resulting from DESI is approaching the lower limit imposed by oscillation experiments. For discussions on the implications for neutrinos, including their total mass and hierarchy, we refer to Refs.~\cite{DESI:2024mwx,Craig:2024tky,Wang:2024hen,Allali:2024aiv,Sen:2024pgb,Green:2024xbb,Elbers:2024sha,Naredo-Tuero:2024sgf,Jiang:2024viw,Reboucas:2024smm,Capozzi:2025wyn,DESI:2025zgx}.} As initially emphasized by the DESI collaboration and corroborated by multiple independent re-analyses~\cite{DESI:2024mwx,DESI:2024kob,DESI:2025zgx,DESI:2025fii}, combining DESI BAO with Planck Cosmic Microwave Background (CMB) data and Type Ia Supernovae (SN) distance moduli measurements -- from either the Pantheon-plus catalog~\cite{Scolnic:2021amr, Brout:2022vxf}, the Union3 compilation~\cite{Rubin:2023ovl}, or the five-year observations of the Dark Energy Survey (DESy5)~\cite{DES:2024tys,DES:2024upw,DES:2024hip} -- yields moderate to strong evidence for a time-evolving DE component, often referred to as Dynamical Dark Energy (DDE). Using the well-known Chevallier-Polarski-Linder (CPL) parameterization~\cite{Chevallier:2000qy, Linder:2002et}, where the DE equation of state (EoS) is given by $w(a) = w_0 + w_a(1 - a)$, the data show a preference for a present-day quintessence-like EoS ($w_0 > -1$) that crossed the phantom barrier ($w_a < 0$). This preference reaches a significance level of about $4.2\sigma$ when combining Planck CMB, DESI BAO and DESy5 SN, while it is reduced to around $2.8\sigma$ when replacing DESy5 with Pantheon-plus~\cite{DESI:2024mwx}.

If confirmed, this result would represent the first concrete evidence of new physics beyond the standard model of cosmology, potentially bringing a significant portion of the theoretical high-energy physics community back to the board. Therefore, somewhat unsurprisingly, following the DESI data release, a multitude of reanalyses and attempts at physical interpretation of this signal have heated up the debate, see, e.g., Refs.~\cite{Cortes:2024lgw,Patel:2024odo,Orchard:2024bve,Liu:2024gfy,Chudaykin:2024gol,Notari:2024rti,Gialamas:2024lyw,Wang:2024hwd,DESI:2024kob,Wang:2024dka,Giare:2024smz,Carloni:2024zpl,Colgain:2024xqj,Bhattacharya:2024hep,Tada:2024znt,Yin:2024hba,Luongo:2024fww,Park:2024jns,Wang:2024sgo,Shlivko:2024llw,Ye:2024ywg,Li:2024qso,Yang:2024kdo,Jiang:2024xnu,Giare:2024gpk,Poulin:2024ken,Pourojaghi:2024bxa,Dinda:2024ktd,Escamilla:2024ahl,Sabogal:2024yha,Ghedini:2024mdu,Wolf:2024eph,Arjona:2024dsr,Wolf:2024stt,Chakraborty:2025syu,Chakraborty:2025syu,Silva:2025hxw,Pan:2025qwy,Pan:2025psn,Ormondroyd:2025iaf,Wolf:2025jed,Anchordoqui:2025fgz,Pang:2025lvh,Colgain:2025nzf,Kessler:2025kju,Nesseris:2025lke,Shah:2025ayl,Santos:2025wiv,Brandenberger:2025hof,Bansal:2025ipo,Lewis:2024cqj,RoyChoudhury:2025dhe,Teixeira:2025czm}. Given its potential impact on our understanding of the Universe, caution is certainly mandatory and thoroughly testing the robustness of these findings is of paramount importance. 

A crucial aspect that has already undergone significant cross-checking is the role played by the parameterization used to describe the DE EoS~\cite{DESI:2024aqx,Hernandez-Almada:2024ost,Pourojaghi:2024tmw,Ramadan:2024kmn,Berghaus:2024kra,Qu:2024lpx,Notari:2024rti,Adolf:2024twn}. In Ref.~\cite{Giare:2024gpk}, we noted that assuming the CPL parameterization is not the primary driver of this preference. 
When combining CMB, DESI BAO, and SN measurements, the same preference for a present-day quintessence-like DDE that crosses the phantom barrier consistently emerges across a wide range of different parameterizations. In this sense, the preference is robust against different models, as subsequently confirmed by the DESI collaboration itself~\cite{DESI:2025fii}.

With the parameterization issue clarified, the next step is to investigate potential observational systematic effects in the datasets involved. Systematic effects in DESI BAO measurements have been a topic of debate. Notably, the DESI BAO measurement at $z = 0.71$ shows approximately $3\sigma$ tension with predictions based on the Planck best-fit $\Lambda$CDM cosmology. This outlier is potentially responsible for part of the DESI signals for new physics, including (part of) the preference for DDE~\cite{Wang:2024pui,Colgain:2024xqj,Naredo-Tuero:2024sgf}. 

Similarly, the impact of SN measurements was recently highlighted in Ref.~\cite{Efstathiou:2024xcq}, where cross-correlating the Pantheon-plus and DESy5 supernova samples suggested evidence for a calibration difference of $\sim 0.04$ mag between low and high redshifts.\footnote{See also Refs.~\cite{Colgain:2022nlb,Colgain:2022rxy,Malekjani:2023ple} for discussions on the consistency of SN measurements across high and low redshifts, as well as Ref.~\cite{Colgain:2024ksa} for other discussions surrounding potential calibration issues in DESy5.} for other discussions surrounding  Correcting for this offset brings the DESy5 sample into closer agreement with Planck’s $\Lambda$CDM cosmology. Since the parameter range favored by the uncorrected DESy5 sample diverges from many other cosmological datasets, it has been suggested that the evidence for DDE might primarily be due to (systematics in) DESy5 SN~\cite{Huang:2024qno,Efstathiou:2024xcq,Notari:2024zmi}. In response to these concerns, however, the DES collaboration in Ref.~\cite{DES:2025tir} showed that the debated $\sim 0.04$ mag offset between Pantheon-plus and DESy5 at low and high redshift is partly attributable to improvements in the modeling of supernova intrinsic scatter and host galaxy properties in DESy5 (accounting for up to $\sim 43\%$ of the offset), and partly ($\sim 38\%$) to a misleading comparison. The latter stems from the fact that different selection functions characterize the DES subsets included in Pantheon-plus and DESy5, leading to differences in individual supernova distance measurements due to distinct bias corrections.

A final missing step (which I will undertake in this work) is to evaluate the role of CMB data in this analysis. Overall, CMB data alone have limited capacity to constrain DDE models due to minimal effects at the epoch of the last scattering surface and the increased number of cosmological parameters~\cite{DiValentino:2017zyq,DiValentino:2019dzu,DiValentino:2022oon}. This challenge, widely acknowledged as geometrical degeneracy, arises because different combinations of late-time parameters can be adjusted so that the acoustic angular scale $\theta_s$ -- defined by the ratio of the comoving sound horizon at recombination to the comoving distance to last scattering -- remains constant if both quantities change proportionally. Consequently, measurements based solely on this geometrical scale cannot strongly constrain parameters affecting the late-time expansion history unless perturbation-level effects and late-time data are also considered. In light of this, one might expect CMB data to have a limited impact on the preference toward DDE in the joint CMB+BAO+SN analyses. Nevertheless, CMB data still play a crucial role in determining the value of the sound horizon at the drag epoch, which serves as a calibrator for BAO measurements, as well as in determining the angular diameter distance from the CMB. Therefore, when combined with BAO and SN data, CMB measurements provide significant additional information and can influence correlations among parameters. Additionally, the DE dynamics, influencing the decay of the gravitational potential, can leave imprints on the late Integrated Sachs-Wolfe (ISW) effect and thus alter the amplitude of the CMB angular power spectrum at large angular scales.

To the best of my knowledge, so far all the analyses surrounding the DESI and SN preference for DDE have primarily (although somewhat unconsciously) relied on temperature and polarization anisotropy spectra measured by the Planck satellite~\cite{Planck:2018vyg,Planck:2019nip}, as well as lensing power spectrum reconstructions from Planck, whether from \texttt{Plik}~\cite{Planck:2018lbu} or the more recent \texttt{NPIPE-PR4}~\cite{Carron:2022eyg} likelihoods.\footnote{See recent Ref.~\cite{RoyChoudhury:2024wri} for updated constraints derived from Planck-PR4 based likelihoods.} The only notable non-Planck contribution to the CMB side of these analyses comes from the Atacama Cosmology Telescope (ACT) lensing likelihood (\texttt{ACT-DR6})~\cite{ACT:2023kun,ACT:2023dou}, often used in conjunction with \texttt{NPIPE-PR4} or \texttt{Plik} lensing. However, this seems a modest usage of the multitude of CMB observations available beyond Planck. Ground-based CMB experiments, such as ACT~\cite{ACT:2020gnv,ACT:2023kun,ACT:2023dou,ACT:2025fju} and the South Pole Telescope (SPT)~\cite{SPT-3G:2021eoc,SPT-3G:2022hvq}, provide measurements of temperature and polarization anisotropies that, while partially overlapping with Planck, extend into higher multipole regions (i.e., smaller angular scales). When considered independently and especially in combination with other large-scale CMB data (such as from WMAP~\cite{WMAP:2012nax}), these experiments can achieve a level of precision on cosmological parameters approaching that of Planck. 

Expanding the analysis and testing whether the preference for DDE is confirmed by CMB measurements other than Planck is certainly feasible and worthwhile for a good number of well-motivated reasons. Firstly, over the years, several mild anomalies have been identified in the Planck CMB measurements, many of which affect parameters important for the late-time dynamics of the Universe. One notable example concerns the higher lensing amplitude inferred from Planck temperature and polarization spectra, captured by the phenomenological parameter $A_{\rm L}$~\cite{Calabrese:2008rt}, which deviates from the baseline value ($A_{\rm L} = 1$) by about $2.8\sigma$~\cite{Planck:2018vyg,DiValentino:2015bja,Renzi:2017cbg,Domenech:2020qay}. This issue can be recast into a well-documented preference for a closed Universe, as discussed in several recent works~\cite{Park:2017xbl,Handley:2019tkm,DiValentino:2019qzk,Efstathiou:2020wem,DiValentino:2020hov,Benisty:2020otr,Vagnozzi:2020rcz,Vagnozzi:2020dfn,DiValentino:2020kpf,Yang:2021hxg,Cao:2021ldv,Dhawan:2021mel,Dinda:2021ffa,Gonzalez:2021ojp,Akarsu:2021max,Cao:2022ugh,Glanville:2022xes,Bel:2022iuf,Yang:2022kho,Stevens:2022evv,Favale:2023lnp}. Secondly, focusing solely on Planck CMB data, one might speculate about a latent preference for a phantom-like DE component~\cite{Planck:2018vyg,Escamilla:2023oce} that is closely related to large-scale temperature and E-modes polarization measurements~\cite{Escamilla:2023oce,Giare:2023ejv,Ben-Dayan:2024uvx}. This latent preference (not confirmed by other CMB experiments) was previously washed out by older BAO measurements from the SDSS surveys~\cite{eBOSS:2020yzd}. However, it is worth checking whether it now sums up with the DESI and SN tendency for a phantom DDE, thereby strengthening its preference. Last but not least, discrepancies often arise between Planck and Planck-independent experiments as reported by different groups~\cite{Lin:2019zdn,Forconi:2021que,Handley:2020hdp,LaPosta:2022llv,DiValentino:2022rdg,DiValentino:2022oon,Giare:2022rvg,Calderon:2023obf,Giare:2023xoc,Giare:2024sdl,Gariazzo:2024sil,Forconi:2025zzu}. Most disagreements involve small-scale CMB measurements from Data Release 4 of the ACT (\texttt{ACT-DR4}), and concern the value of inflationary parameters (most notably the value of the spectral index~\cite{Giare:2022rvg,Giare:2023wzl,Forconi:2025zzu}). While these problems have been largely mitigated in the latest Data Release 6 of the ACT temperature and polarization spectra (\texttt{ACT-DR6}), differences are still observed and can potentially propagate to other cosmological parameters, such as the matter density parameter and the sound horizon~\cite{Giare:2023xoc}, which could have implications for DDE that warrant further investigation.

Driven by these motivations and concerns, I present the results of a comprehensive analysis aimed at clarifying the role of various latest CMB experiments in conjunction with DESI-DR2 BAO and SN measurements. My analysis shows that the preference for DDE is \textit{reduced} when considering CMB experiments beyond Planck. Specifically, when the latest ACT-DR6 data (alone or in combination with WMAP) are analyzed together with DESI-DR2 BAO and Pantheon-plus, there is no strong preference for DDE, and the cosmological constant model essentially falls within or close to the 95 \% confidence level (CL) contours. On the other hand, a preference for DDE is still observed when combining (WMAP+)ACT-DR6 with DESI-DR2 BAO and DESy5, confirming the importance of the latter SN catalogue. However, the statistical significance of this preference is diminished compared to Planck-based results. Similarly, when considering SPT data, I find no convincing preference for DDE in combinations that involve Pantheon-plus, and the preference becomes modest in combinations that involve DESy5. I also contend that the temperature and E-mode polarization anisotropy measurements at large angular scales ($\ell \lesssim 30$) in the Planck data are the subset mostly responsible for reinforcing the shift towards DDE.

The paper is structured as follows: In Sec.~\ref{sec:methodology}, I discuss the methodology and datasets underlying this study. In Sec.~\ref{sec:results}, I present and discuss the most important results. In Sec.~\ref{sec:conclusions} I draw my conclusions.

\section{Methodology}
\label{sec:methodology}

\begin{table}[ht!]
\centering
\renewcommand{\arraystretch}{1.5}
\begin{tabular}{l @{\hspace{2cm}} c}
\toprule
\textbf{Parameter} & \textbf{Prior} \\
\hline \hline
$\Omega_\mathrm{b} h^2$ & $[0.005, 0.1]$ \\
$\Omega_\mathrm{c} h^2$ & $[0.01, 0.99]$ \\
$\log(10^{10} A_\mathrm{s})$ & $[1.61, 3.91]$ \\
$n_\mathrm{s}$ & $[0.8, 1.2]$ \\
$\tau_{\rm reio}$ & $[0.01, 0.8]$ \\
$100\theta_\mathrm{s}$ & $[0.5, 10]$ \\
$w_0$ & $[-3, 1]$ \\
$w_a$ & $[-3, 2]$ \\
\bottomrule
\end{tabular}
\caption{Ranges for the flat prior distributions imposed on the 8 free cosmological parameters. In the analyses involving ACT and SPT data (both on their own and in combination with WMAP), I employ a Gaussian prior $\tau = 0.065 \pm 0.015$ as detailed in Sec.~\ref{sec:methodology}. I examine the impact of this prior on these combinations in \hyperref[appendix:B]{Appendix B}.}
\label{tab-priors}
\end{table}

I focus on a cosmological model described by 8 free parameters: the physical baryon energy density $\Omega_\mathrm{b} h^2$, the physical cold dark matter energy density $\Omega_\mathrm{c} h^2$, the amplitude of the primordial scalar spectrum $A_\mathrm{s}$, its spectral index $n_s$, the optical depth to reionization $\tau_{\rm reio}$, the angular size of the sound horizon $\theta_{\rm{s}}$, and two parameters describing the DE sector — namely, the present-day value of the DE EoS $w_0$ and the parameter describing its dynamical evolution $w_a$. I assume the linear CPL parameterization, $w(a) = w_0 + w_a(1 - a)$, to model its dynamical evolution.

I use the publicly available cosmological code \texttt{CAMB}~\cite{Lewis:1999bs,Howlett:2012mh} to compute the theoretical predictions and explore the posterior distributions of the 8-dimensional parameter space by performing Markov Chain Monte Carlo (MCMC) analyses using the publicly available sampler \texttt{Cobaya}~\cite{Lewis:2002ah,Lewis:2013hha}. For assessing chain convergence, I adopt a convergence diagnostic based on the Gelman-Rubin statistic~\cite{Gelman:1992zz} setting a threshold of $R-1 \lesssim 0.01$. The flat prior ranges within which the parameters are varied are given in Tab.~\ref{tab-priors}.

My reference datasets are the following:
\begin{itemize}
    \item \textbf{DESI:} BAO measurements released by DESI  based on its second year of observations (DR2) of galaxies and quasars~\cite{DESI:2025qqy}, as well as Lyman-$\alpha$~\cite{DESI:2025zpo} tracers. These measurements are summarized in Tab.~IV of Ref.~\cite{DESI:2025zgx}.\footnote{In the analysis, I included the full correlation matrix among different distance measurements using the publicly available likelihoods released for the sampler \texttt{cobaya}.}

    \item \textbf{PP:} Distance moduli measurements of 1701 light curves from 1550 spectroscopically confirmed Type Ia supernovae, sourced from 18 different surveys. These supernovae span a redshift range of $0.01$ to $2.26$, as compiled in the Pantheon-plus sample~\cite{Scolnic:2021amr,Brout:2022vxf}.

    \item \textbf{DESy5:} Distance moduli measurements of 1635 Type Ia supernovae covering the redshift range of $0.10 < z < 1.13$, collected during the full 5 years of the Dark Energy Survey Supernova Program~\cite{DES:2024tys,DES:2024upw,DES:2024hip}. I include 194 low-redshift supernovae in the redshift range of $0.025 < z < 0.1$, which overlap with the Pantheon-plus sample~\cite{Scolnic:2021amr, Brout:2022vxf}.

    \item \textbf{Planck:} CMB temperature and polarization anisotropy power spectra (and their cross-spectra) from the Planck 2018 legacy data release (PR3). Specifically, I use:
    \begin{itemize}
        \item the high-$\ell$ \texttt{Plik} likelihood~\cite{Planck:2018vyg,Planck:2019nip} for the TT spectrum in the multipole range $30 \leq \ell \leq 2508$ and for the TE and EE spectra in the range $30 \leq \ell \leq 1996$; 
        \item the low-$\ell$ \texttt{commander} likelihood~\cite{Planck:2018vyg,Planck:2019nip} for the TT spectrum in the multipole range $2 \leq \ell \leq 29$;
        \item the low-$\ell$ \texttt{SimAll} likelihood~\cite{Planck:2018vyg,Planck:2019nip} for the EE spectrum in the range $2 \leq \ell \leq 29$;
        \item the \texttt{Plik} CMB lensing measurements~\cite{Planck:2018lbu}, reconstructed from the temperature 4-point correlation function.
    \end{itemize}

    \item \textbf{ACT:} latest CMB temperature, polarization and lensing spectra from the ACT collaboration, combined with a Gaussian prior on $\tau = 0.065 \pm 0.015$, as done in Ref.~\cite{ACT:2020gnv}. Specifically, I use:
    \begin{itemize}
        \item the \texttt{ACT-DR6} likelihood~\cite{ACT:2025xdm,ACT:2025fju,ACT:2025tim} for observations of temperature and polarization (TT, TE, EE) spectra measured from the Data Release 6 maps made from ACT data. These cover 19,000 deg$^2$ of sky in bands centered at 98, 150, and 220 GHz, with white noise levels three times lower than Planck in polarization~\cite{ACT:2025fju};
        \item the \texttt{ACT-DR6} lensing likelihood based on the gravitational lensing mass map covering 9400 deg$^2$, reconstructed from CMB measurements by ACT from 2017 to 2021~\cite{ACT:2023kun,ACT:2023dou}. In my analysis, I include only the conservative range of lensing multipoles $40 < \ell < 763$.
    \end{itemize}
    
    \item \textbf{SPT:} CMB temperature and polarization (TT, TE, EE) anisotropy spectra released by the SPT collaboration~\cite{SPT-3G:2021eoc,SPT-3G:2022hvq}, combined with a Gaussian prior on $\tau = 0.065 \pm 0.015$.

    \item \textbf{WMAP:} CMB temperature and polarization data from the WMAP 9-year release~\cite{WMAP:2012nax}. I exclude low-$\ell$ TE data (possibly contaminated by dust) and set the minimum multipole in TE at $\ell = 24$. When combining WMAP with ACT or SPT I always assume a Gaussian prior $\tau = 0.065 \pm 0.015$.
\end{itemize}

Notice that many different subsets of the Planck likelihoods (where multipoles at large or small scales are excluded/included in the analysis) are also considered in my study to thoroughly test the preference towards DDE. To avoid overwhelming this list, I define these reduced Planck datasets in the relevant subsections where they are employed.

\section{Results}
\label{sec:results}

In this section, I present the main results of my analysis. The section is organized as follows:
\begin{itemize}
\item In Sec.~\ref{sec:results_1}, I present the constraints on DDE resulting from different combinations of \textit{independent} CMB experiments in conjunction with DESI BAO measurements and Type-Ia Supernovae. The main aim of this subsection is to demonstrate that the preference towards DDE is reduced in CMB experiments other than Planck.

\item In Sec.~\ref{sec:results_2}, I interpret the strengthened preference for DDE observed in Planck, arguing that it is reinforced by temperature and polarization measurements at large angular scales. To do so, I analyze different combinations of the Planck likelihoods where information at large angular scales is appropriately reduced.

\item In Sec.~\ref{sec:results_3}, I further test my conclusions by examining what happens when combining Planck temperature and polarization measurements at large angular scales (which, as I argued in Sec.~\ref{sec:results_2}, strengthen the preference for DDE) with ACT and SPT small-scale data (which, as I argued in Sec.~\ref{sec:results_1}, reduce the preference). This is the only subsection of my work where I mix different CMB experiments together.
\end{itemize}

\subsection{Constraints from independent CMB experiments}
\label{sec:results_1}

\begin{figure*}[htbp!]
    \centering
    \includegraphics[width=0.7\textwidth]{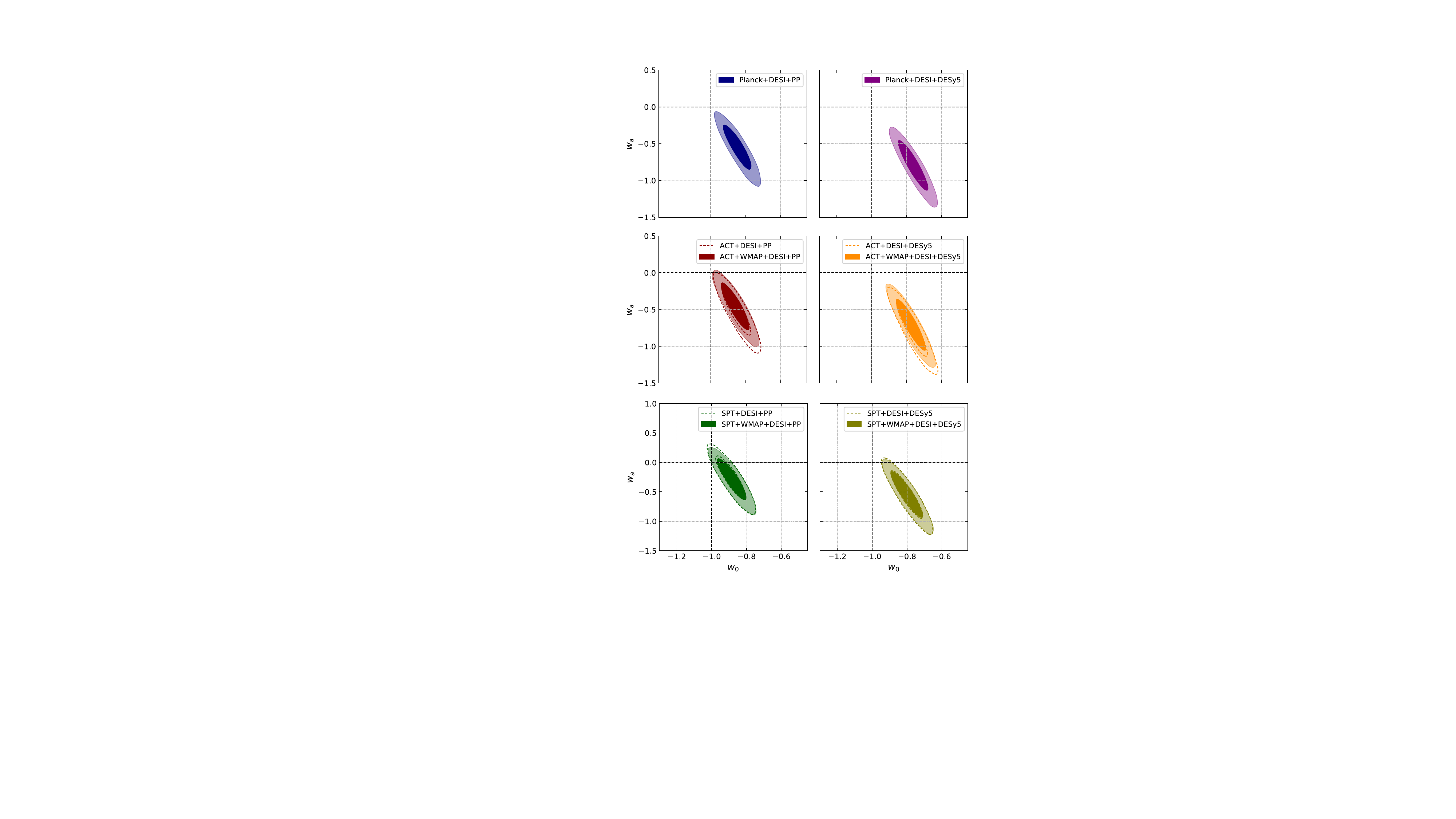}
    \caption{Two-dimensional marginalized contours at 68\% and 95\% CL in the ($w_0$, $w_a$) plane for datasets involving different (independent) CMB experiments. The black dashed lines represent the standard $\Lambda$CDM values, $w_0 = -1$ and $w_a = 0$.}
    \label{fig:1}
\end{figure*}

The constraints on the DE parameters (i.e., the present-day EoS $w_0$ and the parameter $w_a$ quantifying the dynamical behavior) are summarized in Tab.~\ref{tab:results} for the main combinations of data analyzed in this section. For the same combinations of data, the two-dimensional probability contours in the $w_0$-$w_a$ plane are given in Fig.~\ref{fig:1}. Constraints resulting from all other combinations of experiments analyzed in this study, along with the numerical results for all other cosmological parameters, as well as their 1D probability distribution functions and 2D correlations, are provided in \hyperref[appendix:A]{Appendix A}.

\begin{table}[htbp!]
\begin{center}
\renewcommand{\arraystretch}{3}
\resizebox{\columnwidth}{!}{
\begin{tabular}{l | c | c | r}
\hline
\textbf{Datasets} & \boldmath{$w_0$} & \boldmath{$w_a$} & \boldmath{$\Delta \chi^2$} \\ 
\hline\hline
Planck+DESI+PP & 
$ -0.846\pm 0.054$ ($-0.85^{+0.11}_{-0.11}$) & $-0.56^{+0.22}_{-0.19}$ ($-0.56^{+0.39}_{-0.42}$) & $-7.79$ \\
\hline
Planck+DESI+DESy5 & 
$-0.760\pm 0.057$ ($-0.76^{+0.11}_{-0.11}$) & $ -0.80^{+0.24}_{-0.21}$ ($-0.80^{+0.42}_{-0.46}$) & $-19.20$ \\
\hline \hline 

ACT+DESI+PP & 
$-0.851\pm 0.056$ ($-0.85^{+0.11}_{-0.11}$) & $-0.52^{+0.24}_{-0.20}$ ($-0.52^{+0.41}_{-0.46}$) & $-6.14$ \\
\hline
WMAP+ACT+DESI+PP &
$-0.859\pm 0.055$ ($-0.86^{+0.11}_{-0.11}$) & $-0.47^{+0.22}_{-0.20}$ ($-0.47^{+0.40}_{-0.44}$) & $-5.81$ \\
\hline
ACT+DESI+DESy5 & 
$-0.765\pm 0.058$ ($-0.77^{+0.12}_{-0.11}$) & $-0.77^{+0.26}_{-0.23}$ ($-0.77^{+0.46}_{-0.49}$) & $-16.77$ \\
\hline
WMAP+ACT+DESI+DESy5 & 
$-0.774\pm 0.057$ ($-0.77^{+0.11}_{-0.11}$) & $-0.71\pm 0.23$ ($-0.71^{+0.44}_{-0.47}$) & $-15.96$ \\
\hline\hline

SPT+DESI+PP & 
$-0.888\pm 0.057$ ($-0.89^{+0.11}_{-0.11}$) & $-0.27^{+0.26}_{-0.23}$ ($-0.27^{+0.46}_{-0.50}$) & $-3.58$ \\
\hline 
WMAP+SPT+DESI+PP & 
$-0.882\pm 0.056$ ($-0.88^{+0.11}_{-0.11}$) & $-0.29^{+0.25}_{-0.22}$ ($-0.29^{+0.43}_{-0.48}$) & $-4.33$ \\
\hline
SPT+DESI+DESy5 & 
$-0.799\pm 0.061$ ($-0.80^{+0.12}_{-0.12}$) & $-0.55^{+0.28}_{-0.25}$ ($-0.55^{+0.50}_{-0.54}$) & $-14.73$ \\
\hline 
WMAP+SPT+DESI+DESy5 & 
$-0.798\pm 0.061$ ($-0.80^{+0.12}_{-0.11}$) & $-0.55^{+0.28}_{-0.24}$ ($-0.55^{+0.48}_{-0.54}$) & $-14.67$ \\
\hline

\hline \hline
\end{tabular}}
\end{center}
\caption{Constraints on $w_0$ and $w_a$ at 68\% (95\%) CL for the main dataset combinations analyzed in Sec.~\ref{sec:results_1}. $\Delta\chi^2 = \chi^2_{\text{CPL}} - \chi^2_{\Lambda\text{CDM}}$ represents the difference in the best-fit $\chi^2$ with respect to $\Lambda$CDM as obtained within the different datasets. Constraints resulting from other combinations of data or on other cosmological parameters can be found in \hyperref[appendix:A]{Appendix A}.}
\label{tab:results}
\end{table}

I begin by examining Planck-based constraints. These are visually displayed in the top row of Fig.~\ref{fig:1}, with the combination of Planck+DESI+PP shown in the top left panel and Planck+DESI+DESy5 in the top right panel. Unsurprisingly, I find a preference for DDE which is consistent with the findings widely documented in the literature. This preference is already evident in Planck+DESI+PP: I find $w_0 = -0.846 \pm 0.054$ and $w_a = -0.56^{+0.22}_{-0.19}$, both at 68\% CL. As clearly seen from Fig.~\ref{fig:1}, the cosmological constant case falls well outside the 95\% CL contours. However, in line with the narrative reviewed thus far, I confirm that the shift towards DDE is substantially increased when replacing PP with DESy5. Comparing the left and right top panels of Fig.~\ref{fig:1}, it is evident that the marginalized probability contours move towards more negative values of $w_a$ and towards quintessence-like values of $w_0$ that are further away from the cosmological constant. This shift is also evident in the numerical constraints: for Planck+DESI+DESy5 they read $w_0 = -0.760 \pm 0.057$ and $w_a = -0.80^{+0.24}_{-0.21}$ at 68\% CL. An interesting aspect of my re-analysis of Planck CMB data is that I am using the \texttt{Plik} (PR3) temperature and lensing likelihood, whereas the baseline analysis performed by the DESI collaboration used the \texttt{NPIPE-PR4} maps for temperature, polarization, and lensing spectra (the latter in conjunction with \texttt{ACT-DR6} lensing)~\cite{DESI:2025zgx}. On the one hand, here I do not want to mix data from different CMB experiments because I aim to compare results from CMB probes that are as independent as possible (with the exception of Sec.\ref{sec:results_3}, where I briefly examine what happens when combining Planck, ACT, and SPT data together). On the other hand, my analysis shows that using Planck PR3 spectra for temperature, polarization, and lensing spectra leads to results that are largely compatible with those reported by the DESI collaboration. In particular, \texttt{ACT-DR6} lensing and the new \texttt{NPIPE-PR4} maps have a small impact on driving this preference for DDE, with differences remaining around 0.5$\sigma$, consistent with documented results in the literature~\cite{Wang:2024rjd}.

As the next step, I focus on the results based on the ACT temperature, polarization and lensing (\texttt{ACT-DR6}) likelihoods. The marginalized contours for $w_0$ and $w_a$ are shown in the middle row of Fig.~\ref{fig:1}, with ACT(+WMAP)+DESI+PP in the left panel and ACT(+WMAP)+DESI+DESy5 in the right panel. In this case, different remarks are warranted. First and foremost, including WMAP's large-scale temperature and polarization measurements does not significantly impact the constraints on the DE parameters. WMAP measurements provide information around the first acoustic peak and help break degeneracies between cosmological parameters (most notably the degeneracy between the spectral index and the baryon energy density, see also Refs.~\cite{ACT:2020gnv,Giare:2022rvg}). They mainly shift the constraints on these anomalous parameters in ACT back to values inferred by Planck, without substantially improving the constraining power on $w_0$ and $w_a$. More crucial is the role of SN data. The sample used for SN distance moduli measurements now plays a key role in assessing the preference for DDE. If we focus on the combination of ACT(+WMAP)+DESI+PP, it is clear from Fig.~\ref{fig:1} that the cosmological constant case falls close to the 95\% CL contours. This is also confirmed by the numerical results, which for the most constraining combination, ACT+WMAP+DESI+PP, are $w_0 = -0.859\pm 0.055$ and $w_a = -0.47^{+0.22}_{-0.20}$ at 68\% CL. This result is important because it demonstrates that CMB data significantly influence the shift towards DDE. When using ACT+WMAP+DESI+PP, the preference for DDE is notably weaker compared to Planck+DESI+PP, highlighting the crucial role that Planck data play in reinforcing this preference. On the other hand, considering ACT(+WMAP)+DESI+DESy5, I still find a substantial preference for a quintessence-like EoS that crosses into the phantom regime. This is evident both from Fig.~\ref{fig:1} and from the numerical constraints: $w_0 = -0.774\pm 0.057$ and $w_a = -0.71\pm0.23$ at 68\% CL.  This result confirms the findings discussed in Ref.~\cite{Efstathiou:2024xcq} regarding the importance of DESy5 in driving this shift towards DDE. However, despite such a preference being quite significant, it appears reduced compared to Planck+DESI+DESy5. In this regard, an important final remark is that the preference for DDE is not reduced due to larger uncertainties. Comparing the marginalized probability contours obtained for ACT(+WMAP) (middle right panels of Fig.~\ref{fig:1}) with those for Planck (top panels of Fig.~\ref{fig:1}), we see that these combinations of datasets produce essentially the same precision in constraining the parameter space of the model, see also Tab~\ref{tab:results}. Overall, in ACT(+WMAP)-based results, a genuine shift towards $w_a \to 0$ and $w_0 \to -1$ is responsible for the reduced preference. 

The final step I undertake is to analyze the SPT temperature and polarization spectra. In this case, I can already anticipate an overall loss of constraining power due to the significantly larger error bars characterizing the SPT data, as well as the absence of information on the lensing power spectrum. The larger uncertainties obtained in inferring cosmological parameters certainly represent a concern, as they introduce significant correlations between the dark energy parameters $w_0$–$w_a$ and other crucial parameters characterizing the late-time dynamics such as $H_0$ and $\Omega_m$. This is clearly seen in Fig.\ref{fig:4} and Fig.\ref{fig:5} in \hyperref[appendix:A]{Appendix A}. However, despite these caveats, it is important to stress that in this case \textit{no strong preference} for DDE is found in combinations involving Pantheon-plus SN. The results for SPT(+WMAP)+DESI+PP read $w_0 = -0.882 \pm 0.056$ and $w_a = -0.29^{+0.25}_{-0.22}$ at 68\% CL -- in agreement with $\Lambda$CDM within $\sim 2 \sigma$. This is also evident in the left bottom panels of Fig.~\ref{fig:1}. Instead, for SPT(+WMAP)+DESI+DESy5 I get $w_0 = -0.798 \pm 0.061$ and $w_a = -0.55^{+0.28}_{-0.24}$ at 68\% CL. Overall, whether combining SPT(+WMAP)+DESI+PP or SPT(+WMAP)+DESI+DESy5, the resulting marginalized contours in the $w_0$–$w_a$ plane shift significantly closer to the cosmological constant value, even remaining within or at the border of the 95\% CL contours. This outcome reinforces the important role that CMB measurements play in modulating the statistical evidence for DDE.

Overall, my analysis reveals a trend in CMB experiments other than Planck to significantly \textit{reduce} the preference for DDE. This implies that the preference is partially driven by Planck CMB observations. This trend is also evident when examining the $\Delta\chi^2 = \chi^2_{\text{CPL}} - \chi^2_{\Lambda\text{CDM}}$ (last column of Tab~\ref{tab:results}) which represents the difference between the best-fit $\chi^2$ obtained using a CPL parameterization and within the standard $\Lambda$CDM model of cosmology. For combinations of datasets involving Planck, I observe a significant improvement in the fit over $\Lambda$CDM, with $\Delta\chi^2 \sim -7.79$ ($\Delta\chi^2 \sim -19.20$) for Planck+DESI+PP (Planck+DESI+DESy5). However, this improvement is reduced to $\Delta\chi^2 \sim -6.14$ ($-5.81$) for ACT(+WMAP)+DESI+PP and becomes significant again with $\Delta\chi^2 \sim -16.77$ ($-15.96$) for ACT(+WMAP)+DESI+DESy5. Similarly, for SPT, the improvement is very modest, with values ranging from $\Delta\chi^2 \sim -3.58$ to $-4.33$ for SPT(+WMAP)+DESI+PP and deceased to $\Delta\chi^2 \sim -14.7$ for SPT(+WMAP)+DESI+DESy5. Thus, across all results based on different CMB experiments, the $\Delta \chi^2$ follows the same trend as the preference for DDE: the improvement in $\Delta \chi^2$ over $\Lambda$CDM is reduced when the preference for DDE diminishes. This suggests that the shift in the constraints on the parameter space from different CMB data is not merely due to a volume effect but is driven by differences in the fit, further underscoring the significant role that CMB data play in assessing the strength of the preference. Lastly, my analysis reaffirms the crucial role of DESy5 supernovae, as already discussed in the literature. Notably, this is the only dataset that reproduces evidence for DDE in the ACT-based constraints. 

\subsection{Interpreting the Planck preference for DDE}
\label{sec:results_2}

\begin{figure}[htpb!]
    \centering
    \includegraphics[width=0.72\columnwidth]{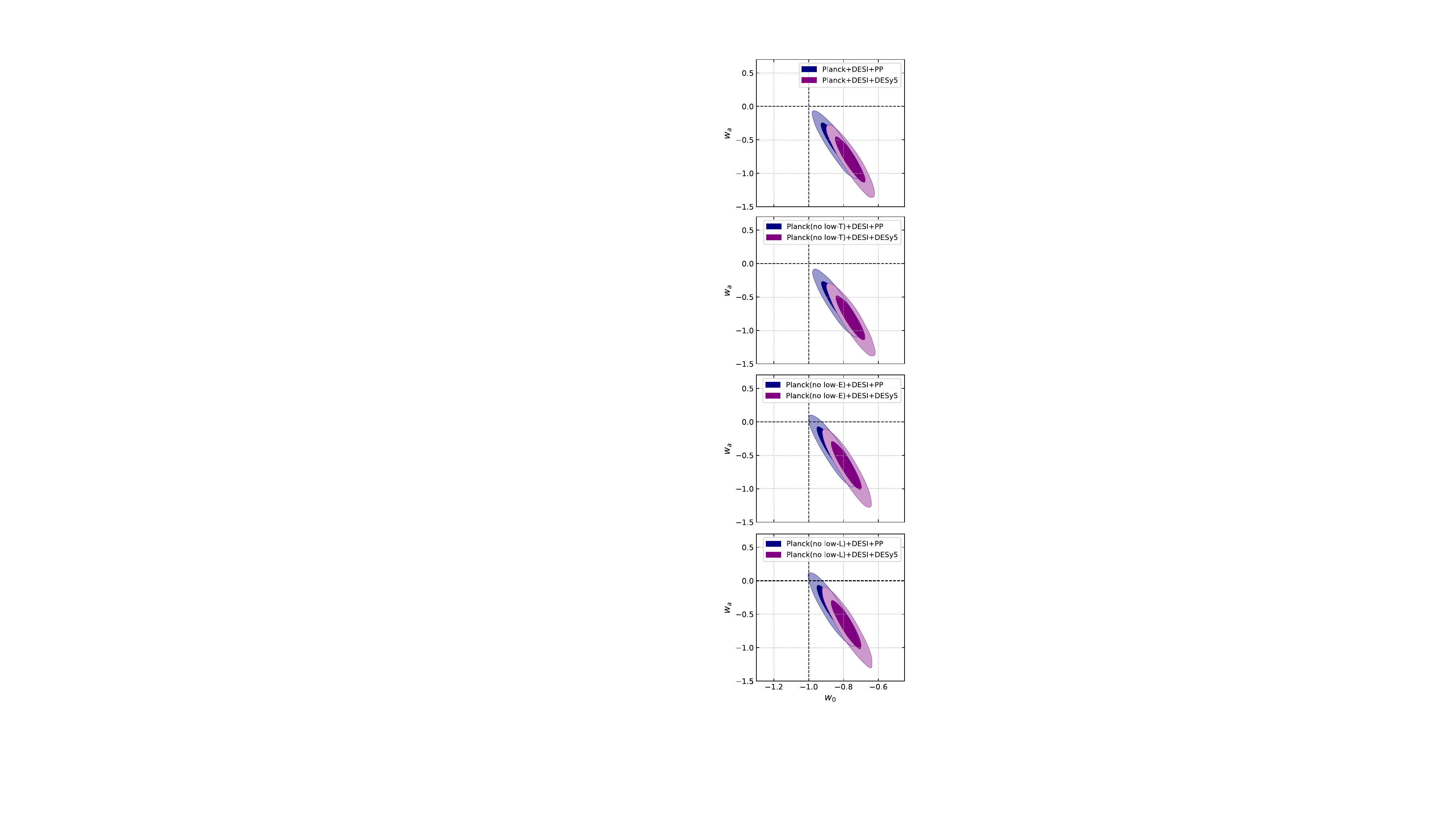}
    \caption{Two-dimensional marginalized contours at 68\% and 95\% CL in the ($w_0$, $w_a$) plane for different combination of the Planck likelihoods. The black dashed lines represent the standard $\Lambda$CDM values, $w_0 = -1$ and $w_a = 0$.}
    \label{fig:2}
\end{figure}

Based on the main conclusion of my analysis (i.e., that CMB experiments other than Planck tend to reduce the preference for DDE), in this section I further investigate the role of Planck data in strengthening this preference and identify which parts of the dataset are mostly responsible for the shift.

A good starting point is to note that the most significant impact on the CMB angular power spectra resulting from the DE dynamics is primarily due to changes in the amplitude of the ISW plateau at very large angular scales. This amplitude is mainly controlled by primordial inflationary parameters (i.e., $A_s$ and $n_s$) and contributions from the late-time ISW effect. The latter is sensitive to DE dynamics, as a phantom (quintessence) DE component can oppose (favor) the decay of gravitational potentials that source the late ISW effect. Additionally, as highlighted several times in the literature, temperature and polarization data at large angular scales ($\ell \le 30$) are crucial for understanding several mild anomalies observed in Planck’s CMB measurements~\cite{Giare:2023ejv}. This includes a latent preference for a phantom-like dark energy component~\cite{Planck:2018vyg,Escamilla:2023oce}, which is closely related to large-scale temperature and E-mode polarization measurements~\cite{Escamilla:2023oce,Giare:2023ejv}, see also \hyperref[appendix:B]{Appendix B} for further discussions. Given these premises, I have a priori reason to hypothesize that part of Planck's tendency to strengthen the preference for DDE could result from large-scale temperature and polarization measurements as well. To corroborate this hypothesis, I performed the following tests: I compare the results obtained within the two baseline datasets involving Planck (Planck+DESI+PP and Planck+DESI+DESy5) with the results obtained for the same combination of data, removing information at low multipoles (i.e., large angular scales). In particular, I consider the following reduced likelihoods for the Planck spectra:
 
\begin{itemize}
\item \textbf{Planck(no low-T)}: In this dataset, I \textit{remove} only the low-$\ell$ \texttt{commander} likelihood for the TT spectrum in the multipole range $2 \leq \ell \leq 29$. Instead, I \textit{keep} all the other likelihoods -- i.e., the high-$\ell$ \texttt{Plik} likelihood for the TT, TE, and EE spectra in the range $30 \leq \ell \lesssim 2500$; the low-$\ell$ \texttt{SimAll} likelihood for the EE spectrum in the range $2 \leq \ell \leq 29$; and the \texttt{Plik} CMB lensing measurements.

\item \textbf{Planck(no low-E)}: In this dataset, I \textit{remove} only the low-$\ell$ \texttt{SimAll} likelihood for the EE spectrum in the range $2 \leq \ell \leq 29$, \textit{keeping} all the other likelihoods -- i.e., the low-$\ell$ \texttt{commander} likelihood for the TT in $2 \leq \ell \leq 29$; the high-$\ell$ \texttt{Plik} likelihood for the TT, TE, and EE spectra in the range $30 \leq \ell \lesssim 2500$; and the \texttt{Plik} CMB lensing measurements.

\item \textbf{Planck(no low-L)}: In this dataset, I \textit{remove} both the low-$\ell$ \texttt{commander} likelihood for the TT spectrum and the low-$\ell$ \texttt{SimAll} likelihood for the EE spectrum. Therefore, I \textit{do not} keep any data in the range $2 \leq \ell \leq 29$ for either TT or EE. Instead, I \textit{keep} the high-$\ell$ \texttt{Plik} likelihood for the TT, TE, and EE spectra and the \texttt{Plik} CMB lensing measurements.
\end{itemize}

The numerical results on the cosmological parameters obtained by reducing information in the Planck likelihoods are presented in \hyperref[appendix:A]{Appendix A} (Tabs.~\ref{Tab:Results_P18_PP} and \ref{Tab:Results_P18_DESy5}) and summarized in Fig.~\ref{fig:2}. Referring to the figure, we can compare step by step how the constraints obtained from the different combinations of datasets change by cutting information at large angular scales. 

In the top panel of the figure, for reference, I plot the results obtained within the baseline combination of data (Planck+DESI+PP and Planck+DESI+DESy5) where all information at large scales is retained. In the second panel from the top, I show the two-dimensional probability contours obtained for Planck(no low-T)+DESI+PP and Planck(no low-T)+DESI+DESy5 -- i.e., removing the low-$\ell$ \texttt{commander} likelihood for the TT spectrum. As seen in the figure, the constraints slightly shift toward the cosmological constant value for both datasets, but the effect is very modest. I obtain $w_0 =  -0.844\pm 0.054$ and $w_a = -0.58^{+0.22}_{-0.19}$ for Planck(no low-T)+DESI+PP, and $w_0 = -0.759\pm 0.056$ and $w_a = -0.81^{+0.23}_{-0.21}$ for Planck(no low-T)+DESI+DESy5 (all at 68\% CL), underscoring that the preference for DDE reduces, but remains solid. A more significant effect is seen in the third panel from the top. In this case, I am removing only the low-$\ell$ \texttt{SimAll} likelihood for the EE spectrum. As I and other collaborators argued in Refs.~\cite{Escamilla:2023oce,Giare:2023ejv}, large-scale E-mode polarization measurements are of crucial importance for the constraints on the DE EoS. In this case, remarkably, the cosmological constant value falls within the 95\% CL contours for Planck(no low-E)+DESI+PP, where I obtain $w_0 = -0.867\pm 0.055$ and $w_a = -0.41^{+0.24}_{-0.20}$ at 68\% CL. I would like to underline the significant shift in the central value of $w_a$. A preference for a present-day quintessence-like EoS that turned phantom in the past remains substantial for Planck(no low-E)+DESI+DESy5: $w_0 = -0.783\pm 0.058$ and $w_a = -0.66^{+0.26}_{-0.22}$ at 68\% CL. However, this preference is clearly reduced compared to the baseline case, again due to a shift in the value of $w_a$. 
When I remove any Planck temperature and polarization measurements at $\ell \lesssim 30$, the preference for DDE is further reduced. Referring to the bottom panel of Fig.~\ref{fig:2}, we see that the cosmological constant value falls within the 95\% CL regions for Planck(no low-L)+DESI+PP, where I obtain $w_0 = -0.868\pm 0.056$ and $w_a = -0.41^{+0.23}_{-0.21}$ at 68\% CL. Conversely, the combination Planck(no low-L)+DESI+DESy5 still shows a significant preference for a dynamical EoS, yielding $w_0 =  -0.782\pm 0.058$ and $w_a = -0.67^{+0.25}_{-0.22}$ at 68\% CL. That said, when removing low-$\ell$ data, the shift in the value of $w_a$ toward less negative values remains.

Given these results, there is solid ground to conclude that Planck temperature and polarization measurements at large angular scales $\ell \lesssim 30$ (especially E-mode polarization measurements, see \hyperref[appendix:B]{Appendix B}.) further strengthen the preference for DDE observed in DESI BAO and Type-Ia SN. Indeed, by removing these, the preference is reduced, and Planck(no low-L)-based constraints behave similarly to ACT-based constraints. 

Last but not least, by comparing the best fit to the CMB angular power spectra obtained within the baseline $\Lambda$CDM cosmology and the CPL model for DDE across different combinations of data, I observe a significant difference in the fit of Planck data around the first acoustic peak in the TT spectrum between the two models for Planck+DESI+DESy5. These differences primarily arise due to shifts in the values of other cosmological parameters within $\Lambda$CDM, which for Planck+DESI+DESy5 leads to a best-fit spectrum that appears in notable tension with both Planck data and the Planck-only best-fit cosmology. This tension is mitigated when allowing $w_0$ and $w_a$ to vary. However, these differences are much reduced in the Planck+DESI+PP best-fit results. Additionally, my analysis reveals that for (WMAP+)ACT+DESI+DESy5, the preference for DDE is found even when including WMAP data around the first peak, or without CMB data at all. Therefore, while Planck data around the first acoustic peak in the TT spectrum might not play a primary role in strengthening the preference for DDE, they significantly contribute to the improvement in $\Delta \chi^2$ over $\Lambda$CDM observed in Table~\ref{tab:results} for Planck+DESI+DESy5.

\subsection{Joint Constraints from Planck, ACT and SPT}
\label{sec:results_3}

To further validate my conclusion concerning the importance of large-scale Planck data, I perform a final test. Namely, despite the main aim of my analysis being to compare the results on DDE from independent CMB experiments, in this section I briefly check what happens when combining ACT and SPT data at small angular scales (which I argued reduce the DESI and SN preference for DDE) with Planck data at large angular scales (which I argued strengthen such a preference). To perform this test, I consider the following reduced Planck likelihood:
\begin{itemize}
    \item \textbf{Planck-650:} I \textit{keep} the low-$\ell$ \texttt{commander} likelihood for the TT spectrum and the low-$\ell$ \texttt{SimAll} likelihood for the EE spectrum (both at $\ell \lesssim 30$); I also \textit{keep} the the high-$\ell$ \texttt{Plik} likelihood for the TT, TE, and EE spectra, focusing only on the \textit{reduced} multipole range $30 \leq \ell \lesssim 650$. I \textit{do not} include the \texttt{Plik} lensing likelihood.
\end{itemize}
Note that I retain the Planck temperature and polarization measurements at large angular scales while cutting the high-multipole measurements at $\ell = 650$ to avoid double-counting the same portion of the sky already measured by ACT and SPT data.\footnote{This is the same approach followed by the ACT collaboration, see e.g., Ref~\cite{ACT:2020gnv,ACT:2025fju}.} This reduced information in the Planck dataset allows me to combine this likelihood with the full temperature, polarization, and lensing \texttt{ACT-DR6} likelihoods, as well as with the full SPT temperature and polarization likelihoods.

\begin{figure}[ht!]
    \centering
    \includegraphics[width=0.72\columnwidth]{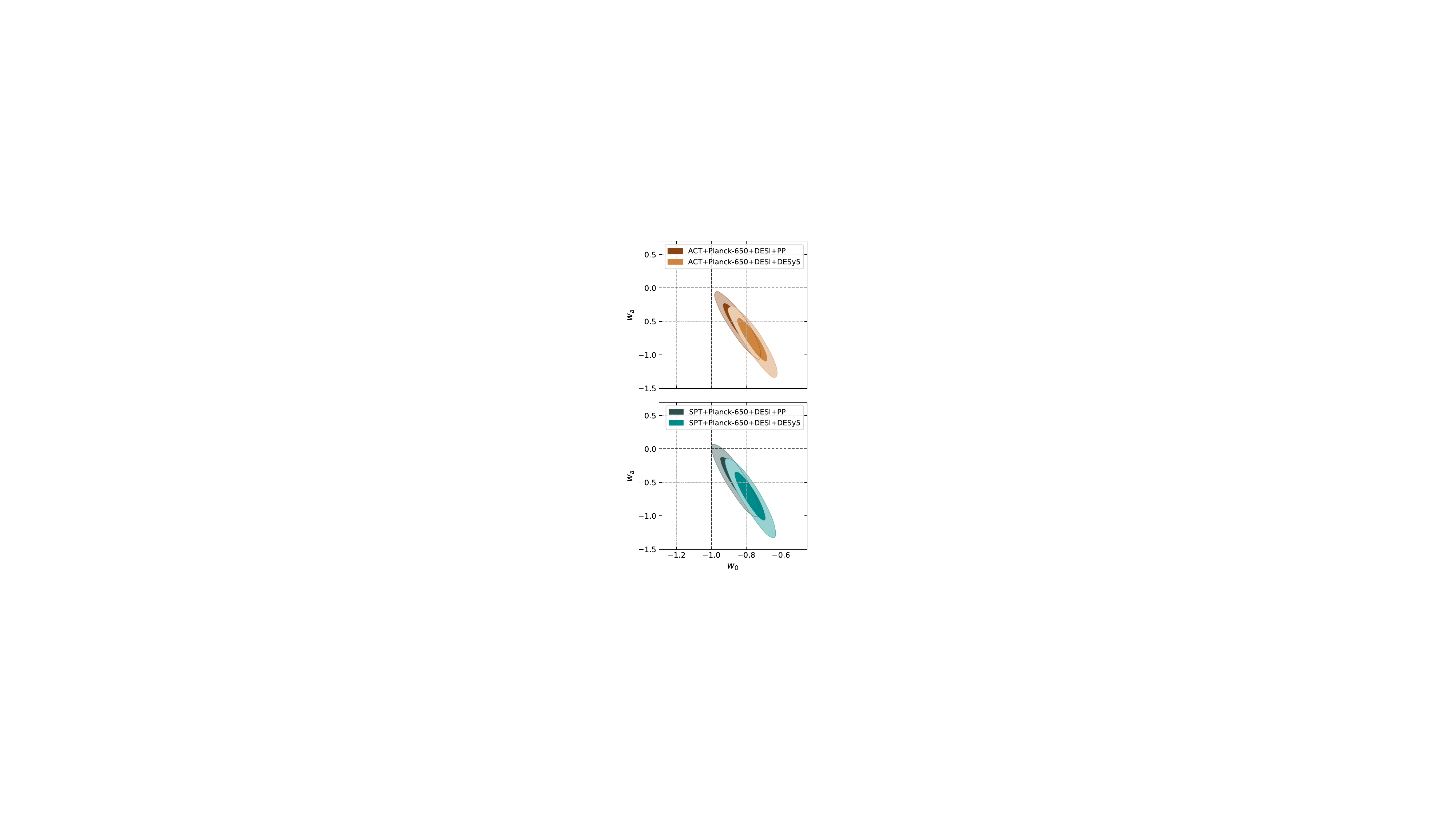}
    \caption{Two-dimensional marginalized contours at 68\% and 95\% CL in the ($w_0$, $w_a$) plane from different mixed combination of the Planck-650, ACT and SPT likelihoods. The black dashed lines represent the standard $\Lambda$CDM values, $w_0 = -1$ and $w_a = 0$.}
    \label{fig:3}
\end{figure}

The numerical constraints on the cosmological parameters obtained by combining Planck-650, ACT, SPT, DESI BAO measurements, and type Ia SN from either the PP or DESy5 catalogues are presented in \hyperref[appendix:A]{Appendix A} (last columns of Tabs.~\ref{Tab:Results_ACT_PP}, \ref{Tab:Results_ACT_DESy5}, \ref{Tab:Results_SPT_PP}, \ref{Tab:Results_SPT_DESy5}) and summarized in Fig.~\ref{fig:3} (see also Fig.~\ref{fig:6} and Fig.~\ref{fig:7} in \hyperref[appendix:A]{Appendix A}) . As seen in the top panel of the figure, when considering the combinations ACT+Planck-650+DESI+PP or ACT+Planck-650+DESI+DESy5, the preference for DDE is well confirmed in both cases. In the former case, the constraints on the DE parameters are $w_0=-0.845\pm 0.054$ and $w_a=-0.55^{+0.21}_{-0.19}$, while for the latter dataset, I get $w_0= -0.759\pm 0.057$ and $w_a= -0.79^{+0.23}_{-0.20}$ (all at 68\% CL). Therefore, the preference is strengthened compared to the same data combinations where Planck-650 was replaced by WMAP, while appearing very similar to Planck+DESI+PP or Planck+DESI+DESy5, respectively. This confirms the overall conclusion that part of the preference for DDE is driven by large-scale temperature and polarization measurements at $\ell \lesssim 30$ (specifically E-mode polarization measurements, see also \hyperref[appendix:B]{Appendix B}), as this portion of information is actually missing in WMAP. On the other hand, from the bottom panel of Fig.~\ref{fig:3}, no clear preference for DDE is observed in the SPT+PP-based results. In the case of SPT+Planck-650+DESI+PP, the cosmological constant lies within the 95\% CL, with constraints $w_0 = -0.860 \pm 0.056$ and $w_a = -0.46^{+0.23}_{-0.21}$ at 68\% CL. For SPT+Planck-650+DESI+DESy5, a preference for dynamical dark energy is again observed (although weakened), with $w_0 = -0.774 \pm 0.059$ and $w_a = -0.71^{+0.26}_{-0.22}$ at 68\% CL.

In conclusion, the joint analysis of Planck, ACT, and SPT data supports the interpretation that part of the preference for DDE originates from the large-scale temperature and polarization measurements from Planck. When these data are reintroduced in the analysis of other CMB experiments, the preference for DDE is significantly strengthened. As a final remark, I note that this is not due to a volume effect: whether mixing different CMB experiments as done in this section, considering the independent combinations as in Sec.~\ref{sec:results_1}, or even cutting information as in Sec.~\ref{sec:results_2}, the uncertainties on $w_0$ and $w_a$ remain essentially unchanged. However, whenever I include large-scale measurements from Planck in the analysis, I observe a genuine shift in their central values towards the portion of parameter space that favors a present-day quintessence-like EoS, which crossed the phantom divide.

\section{Conclusion}
\label{sec:conclusions}

The DESI collaboration has recently released BAO measurements based on its second year of observations that, when combined with Planck CMB data and Type Ia Supernovae from either the Pantheon-plus or DESy5 surveys, produce a preference for Dynamical Dark Energy at a significance ranging between 2.8 and 4.2 standard deviations. In this work, I investigate the role of Planck CMB measurements in favoring this preference and test whether and to what extent the shift toward evolving Dark Energy is supported by other CMB experiments.

I assume a linear Chevallier-Polarski-Linde parameterization to describe the evolution of the Dark Energy equation of state and compare the observational constraints obtained from various combinations of CMB experiments. 

Specifically, in Sec.~\ref{sec:results_1}, I first focus \textit{Planck-independent} combinations of CMB experiments analyzing the latest Atacama Cosmology Telescope and South Pole Telescope measurements of temperature, polarization, and lensing spectra at small scales, eventually combined with WMAP 9-year observations to include information at large angular scales. My analysis reveals that CMB experiments other than Planck tend to \textit{reduce} the preference for evolving Dark Energy. This conclusion is evident from the numerical results on the dark energy parameters $w_0$ and $w_a$ summarized in Tab.~\ref{tab:results}, as well as from Fig.~\ref{fig:1}, which shows the marginalized probability contours at 68\% and 95\% confidence levels in the $w_0$-$w_a$ plane. As seen in Fig.~\ref{fig:1} and Tab.~\ref{tab:results}, when the Atacama Cosmology Telescope temperature, polarization, and lensing data (either alone or combined with WMAP 9-year observations) are used alongside DESI BAO and supernova measurements, they achieve a constraining power comparable to that of Planck. However, the preference for Dynamical Dark Energy is significantly reduced when focusing on the combination involving ACT(+WMAP), DESI BAO, and Pantheon-plus supernovae. In this case, the cosmological constant falls close to the 95\% confidence level contours. A preference for Dynamical Dark Energy reemerges when Pantheon-plus is replaced with DESy5 supernova data, though it remains less pronounced compared to results based on Planck CMB measurements. This reduction in the preference for evolving Dark Energy is confirmed when comparing the improvement in the fit over $\Lambda$CDM. This information is detailed in the last column of Tab.~\ref{tab:results}, where I report the best-fit $\Delta\chi^2 = \chi^2_{\text{CPL}} - \chi^2_{\Lambda\text{CDM}}$ for each combination of datasets analyzed in my work. The improvement in the fit is always reduced in ACT-based results compared to those based on Planck data. The interpretation of the results becomes somewhat more challenging when extending the analysis to the SPT data. In this case, the larger uncertainties obtained in inferring cosmological parameters present both a challenge and a concern, introducing significant correlations among the dark energy parameters $w_0$ and $w_a$ and other crucial parameters characterizing the late time dynamics such as $H_0$ and $\Omega_m$, see also Fig.~\ref{fig:4} and Fig.~\ref{fig:5} in \hyperref[appendix:A]{Appendix A}. That said, as seen in Fig.~\ref{fig:1}, the preference for Dynamical Dark Energy is essentially lost when SPT is combined with Pantheon-plus SN and is significancy reduced for combinations involving DESy5. This is also reflected in the behavior of $\Delta\chi^2$.

Given these results, as a second step, I further investigate the role of Planck data. In Sec.~\ref{sec:results_2}, I argue that the subset of Planck measurements strengthening this preference are the temperature and polarization measurements at large angular scales $\ell \lesssim 30$, especially the E-mode polarization measurements. This is somewhat expected, as the most significant impact of Dark Energy dynamics on the CMB angular power spectrum typically arises from changes in the amplitude of the Integrated Sachs-Wolfe plateau, which affect these scales in the TT spectrum, or from correlations between Dark Energy parameters and other cosmological parameters such as $\tau$, $A_s$, and $\Omega_m$, which can also be sensitive to arge scale measurements (see, e.g., Ref.~\cite{Giare:2023ejv} and \hyperref[appendix:B]{Appendix B}).  As seen in Fig.~\ref{fig:2}, cutting down information at $\ell \lesssim 30$ from the Planck likelihoods, the cosmological constant value falls within the 95\% CL contours for the constraints based on Pantheon-plus supernovae due to a significant shift in the central value of $w_a$. On the other hand, a preference for evolving Dark Energy persists in the results based on DESy5 supernovae, although to a lesser extent. To further validate my conclusions regarding the importance of Planck's temperature and polarization measurements at large angular scales, in Sec.~\ref{sec:results_3}, I perform a joint analysis of Planck large-scale temperature and polarization data (which I argued strengthen the preference) with the Atacama Cosmology Telescope and South Pole Telescope small-scale measurements (which, when analyzed on their own or in combination with WMAP, reduce the preference). The joint analysis supports the interpretation that part of the shift toward Dynamical Dark Energy is reinforced by Planck's large-scale temperature and polarization anisotropy measurements. Indeed, as seen in Fig.~\ref{fig:3}, when Planck's large-scale data are analyzed in combination with the Atacama Cosmology Telescope likelihoods, the preference for evolving Dark Energy is strengthened compared to the case without Planck. Interestingly, no strong preference is observed when combining Planck’s large-scale data with the South Pole Telescope likelihoods alongside Pantheon-plus SN.

In conclusion:
\begin{itemize}
\item The Planck data, particularly large-scale temperature and E-mode polarization measurements, play a significant role in strengthening the preference for Dynamical Dark Energy reported by the DESI collaboration.
\item The preference diminishes when considering Planck-independent CMB observations or excluding Planck large-scale data. In these cases, no strong preference is observed in combinations of data involving DESI and Pantheon-plus supernovae, and the cosmological constant value consistently falls within (or close to) the 95\% confidence level.
\item The preference is consistently strengthened when including the DESy5 supernova data compared to Pantheon-plus-based results. In fact, this is the only supernova sample that produces a convincing preference in combinations involving Atacama Cosmology Telescope data (without Planck) or when Planck large-scale data are removed.
\item No convincing preference is seen in any combinations involving South Pole Telescope data and Pantheon-plus, while it is reduced in combinations involving South Pole Telescope data and DESy5 SN (both including or excluding Planck/WMAP large-scale temperature and polarization measurements).
\end{itemize}

When it comes to interpreting these findings, they present a double-edged sword. On the one hand, one could argue that retaining \textit{all} information from large to small scales across different CMB experiments (see, e.g., Planck-650+ACT-based results) still hints at a preference for Dynamical Dark Energy, which could be interpreted as a resilience of the signal~\cite{Giare:2025pzu}. On the other hand, it remains true that this preference is primarily reinforced by Planck’s large-scale data and the DESy5 supernovae. The former have consistently driven mild anomalies and deviations from $\Lambda$CDM predictions~\cite{Escamilla:2023oce,Giare:2023ejv,Ben-Dayan:2024uvx}, while the latter have recently been subject to debate due to potential calibration issues~\cite{Efstathiou:2024xcq}. Additionally, the presence of unexplained outliers in the preliminary DESI BAO measurements could also contribute to the overall shift toward Dynamical Dark Energy~\cite{Wang:2024pui,Colgain:2024xqj,Naredo-Tuero:2024sgf}. Wrapping everything up, for those of us adopting a more conservative stance, these findings may warrant additional caution in assessing the robustness of the DESI BAO and supernova preference for Dynamical Dark Energy. Further tests and future data -- including upcoming CMB observations and future BAO data releases from DESI -- will be invaluable in providing a definitive answer.

\begin{acknowledgments}
\noindent I thank Eleonora Di Valentino, Alessandro Melchiorri, and Elsa M. Teixeira for interesting discussions and suggestions on a preliminary version of this work. I also thank the referee for the valuable suggestions that led to the additional tests presented in \hyperref[appendix:B]{Appendix B}, which offer further insight into the interpretation of the results. I acknowledge support from the Lancaster–Sheffield Consortium for Fundamental Physics through the Science and Technology Facilities Council (STFC) grant ST/X000621/1. This article is based upon work from COST Action CA21136 Addressing observational tensions in cosmology with systematics and fundamental physics (CosmoVerse) supported by COST (European Cooperation in Science and Technology). I acknowledge IT Services at The University of Sheffield for the provision of services for High Performance Computing.
\end{acknowledgments}

\bibliographystyle{apsrev4-1}
\bibliography{main.bib}
\widetext
\appendix
\clearpage
\section{Numerical constraints on cosmological parameters and contour plots}
\label{appendix:A}

\begin{table*}[h]
\begin{center}
\renewcommand{\arraystretch}{1.5}
\resizebox{\textwidth}{!}{
\begin{tabular}{l c c c c c c c c c c c c c c c }
\hline
\textbf{Parameter} & \textbf{ Planck+DESI+PP } & \textbf{Planck(no low-T)+DESI+PP } & \textbf{ Planck(no low-E)+DESI+PP } & \textbf{ Planck(no low-L)+DESI+PP } \\ 
\hline\hline

$ w_0  $ & $  -0.846\pm 0.054\, ( -0.85^{+0.11}_{-0.11} ) $ & $  -0.844\pm 0.054\, ( -0.84^{+0.11}_{-0.11} ) $ & $  -0.867\pm 0.055\, ( -0.87^{+0.11}_{-0.11} ) $ & $  -0.868\pm 0.056\, ( -0.87^{+0.11}_{-0.11} ) $ \\ 
$ w_{a}  $ & $  -0.56^{+0.22}_{-0.19}\, ( -0.56^{+0.39}_{-0.42} ) $ & $  -0.58^{+0.22}_{-0.19}\, ( -0.58^{+0.39}_{-0.42} ) $ & $  -0.41^{+0.24}_{-0.20}\, ( -0.41^{+0.42}_{-0.43} ) $ & $  -0.41^{+0.23}_{-0.21}\, ( -0.41^{+0.42}_{-0.45} ) $ \\ 
$ \Omega_\mathrm{b} h^2  $ & $  0.02245\pm 0.00013\, ( 0.02245^{+0.00026}_{-0.00026} ) $ & $  0.02245\pm 0.00014\, ( 0.02245^{+0.00027}_{-0.00027} ) $ & $  0.02253\pm 0.00014\, ( 0.02253^{+0.00029}_{-0.00028} ) $ & $  0.02254\pm 0.00015\, ( 0.02254^{+0.00030}_{-0.00029} ) $ \\ 
$ \Omega_\mathrm{c} h^2  $ & $  0.11901\pm 0.00087\, ( 0.1190^{+0.0017}_{-0.0017} ) $ & $  0.11917\pm 0.00090\, ( 0.1192^{+0.0017}_{-0.0018} ) $ & $  0.1180\pm 0.0011\, ( 0.1180^{+0.0022}_{-0.0022} ) $ & $  0.1181\pm 0.0011\, ( 0.1181^{+0.0022}_{-0.0023} ) $ \\ 
$ \tau_\mathrm{reio}  $ & $  0.0555\pm 0.0072\, ( 0.055^{+0.015}_{-0.014} ) $ & $  0.0553\pm 0.0074\, ( 0.055^{+0.015}_{-0.014} ) $ & $  0.078\pm 0.017\, ( 0.078^{+0.033}_{-0.033} ) $ & $  0.080\pm 0.017\, ( 0.080^{+0.033}_{-0.033} ) $ \\ 
$ H_0  $ & $  67.65\pm 0.60\, ( 67.6^{+1.2}_{-1.2} ) $ & $  67.65\pm 0.59\, ( 67.7^{+1.2}_{-1.1} ) $ & $  67.60\pm 0.59\, ( 67.6^{+1.2}_{-1.2} ) $ & $  67.60\pm 0.59\, ( 67.6^{+1.2}_{-1.1} ) $ \\ 
$ n_\mathrm{s}  $ & $  0.9679\pm 0.0036\, ( 0.9679^{+0.0071}_{-0.0071} ) $ & $  0.9664\pm 0.0037\, ( 0.9664^{+0.0072}_{-0.0072} ) $ & $  0.9710\pm 0.0043\, ( 0.9710^{+0.0086}_{-0.0082} ) $ & $  0.9699\pm 0.0043\, ( 0.9699^{+0.0087}_{-0.0085} ) $ \\ 
$ \log(10^{10} A_\mathrm{s})  $ & $  3.045\pm 0.014\, ( 3.045^{+0.028}_{-0.028} ) $ & $  3.045\pm 0.015\, ( 3.045^{+0.029}_{-0.028} ) $ & $  3.087\pm 0.031\, ( 3.087^{+0.060}_{-0.061} ) $ & $  3.090\pm 0.032\, ( 3.090^{+0.061}_{-0.062} ) $ \\ 

\hline \hline
\end{tabular} }
\end{center}
\caption{Results at 68\% (95\%) CL on cosmological parameters obtained considering (different subsets of) Planck CMB likelihoods in combination with DESI BAO and Pantheon-plus SN distance moduli measurements.}
\label{Tab:Results_P18_PP}
\end{table*}

\begin{table*}[h]
\begin{center}
\renewcommand{\arraystretch}{1.5}
\resizebox{\textwidth}{!}{
\begin{tabular}{l c c c c c c c c c c c c c c c }
\hline
\textbf{Parameter} & \textbf{ Planck+DESI+DESy5 } & \textbf{Planck(no low-T)+DESI+DESy5 } & \textbf{ Planck(no low-E)+DESI+DESy5 } & \textbf{ Planck(no low-L)+DESI+DESy5 }\\ 
\hline\hline

$ w_0  $ & $  -0.760\pm 0.057\, ( -0.76^{+0.11}_{-0.11} ) $ & $  -0.759\pm 0.056\, ( -0.76^{+0.11}_{-0.11} ) $ & $  -0.783\pm 0.058\, ( -0.78^{+0.12}_{-0.11} ) $ & $  -0.782\pm 0.058\, ( -0.78^{+0.12}_{-0.11} ) $ \\ 
$ w_{a}  $ & $  -0.80\pm 0.22\, ( -0.80^{+0.42}_{-0.45} ) $ & $  -0.81^{+0.23}_{-0.20}\, ( -0.81^{+0.42}_{-0.45} ) $ & $  -0.66^{+0.26}_{-0.22}\, ( -0.66^{+0.46}_{-0.48} ) $ & $  -0.67^{+0.25}_{-0.22}\, ( -0.67^{+0.46}_{-0.50} ) $ \\ 
$ \Omega_\mathrm{b} h^2  $ & $  0.02244\pm 0.00013\, ( 0.02244^{+0.00027}_{-0.00026} ) $ & $  0.02244\pm 0.00013\, ( 0.02244^{+0.00026}_{-0.00026} ) $ & $  0.02251\pm 0.00015\, ( 0.02251^{+0.00028}_{-0.00028} ) $ & $  0.02251\pm 0.00015\, ( 0.02251^{+0.00029}_{-0.00028} ) $ \\ 
$ \Omega_\mathrm{c} h^2  $ & $  0.11925\pm 0.00090\, ( 0.1192^{+0.0017}_{-0.0018} ) $ & $  0.11936\pm 0.00090\, ( 0.1194^{+0.0017}_{-0.0018} ) $ & $  0.1184\pm 0.0011\, ( 0.1184^{+0.0022}_{-0.0023} ) $ & $  0.1184\pm 0.0011\, ( 0.1184^{+0.0022}_{-0.0022} ) $ \\ 
$ \tau_\mathrm{reio}  $ & $  0.0544\pm 0.0072\, ( 0.054^{+0.015}_{-0.014} ) $ & $  0.0546\pm 0.0072\, ( 0.055^{+0.014}_{-0.014} ) $ & $  0.074\pm 0.017\, ( 0.074^{+0.033}_{-0.033} ) $ & $  0.076\pm 0.017\, ( 0.076^{+0.033}_{-0.033} ) $ \\ 
$ H_0  $ & $  66.88\pm 0.56\, ( 66.9^{+1.1}_{-1.1} ) $ & $  66.88\pm 0.56\, ( 66.9^{+1.1}_{-1.1} ) $ & $  66.87\pm 0.56\, ( 66.9^{+1.1}_{-1.1} ) $ & $  66.86\pm 0.56\, ( 66.9^{+1.1}_{-1.1} ) $ \\ 
$ n_\mathrm{s}  $ & $  0.9673\pm 0.0036\, ( 0.9673^{+0.0070}_{-0.0071} ) $ & $  0.9660\pm 0.0037\, ( 0.9660^{+0.0073}_{-0.0071} ) $ & $  0.9699\pm 0.0043\, ( 0.9699^{+0.0085}_{-0.0085} ) $ & $  0.9689\pm 0.0044\, ( 0.9689^{+0.0086}_{-0.0085} ) $ \\ 
$ \log(10^{10} A_\mathrm{s})  $ & $  3.043\pm 0.014\, ( 3.043^{+0.029}_{-0.028} ) $ & $  3.044\pm 0.014\, ( 3.044^{+0.028}_{-0.028} ) $ & $  3.079\pm 0.031\, ( 3.079^{+0.061}_{-0.061} ) $ & $  3.082\pm 0.031\, ( 3.082^{+0.061}_{-0.060} ) $ \\ 

\hline \hline
\end{tabular} }
\end{center}
\caption{Results at 68\% (95\%) CL on cosmological parameters obtained considering (different subsets of) Planck CMB likelihoods in combination with DESI BAO and DESy5 SN distance moduli measurements.}
\label{Tab:Results_P18_DESy5}
\end{table*}

\begin{table*}[h]
\begin{center}
\renewcommand{\arraystretch}{1.5}
\resizebox{\textwidth}{!}{
\begin{tabular}{l c c c c c c c c c c c c c c c }
\hline
\textbf{Parameter} & \textbf{ ACT+DESI+PP } & \textbf{ ACT+WMAP+DESI+PP } & \textbf{ ACT+Planck-650+DESI+PP } \\ 
\hline\hline

$ w_0 $ & $  -0.852^{+0.052}_{-0.058}\, ( -0.85^{+0.11}_{-0.11} ) $ & $  -0.860\pm 0.053\, ( -0.86^{+0.11}_{-0.10} ) $ & $  -0.845\pm 0.054\, ( -0.85^{+0.11}_{-0.11} ) $ \\ 
$ w_{a}  $ & $  -0.52^{+0.24}_{-0.20}\, ( -0.52^{+0.41}_{-0.47} ) $ & $  -0.46^{+0.22}_{-0.20}\, ( -0.46^{+0.38}_{-0.43} ) $ & $  -0.55^{+0.21}_{-0.19}\, ( -0.55^{+0.38}_{-0.41} ) $ \\ 
$ \Omega_\mathrm{b} h^2  $ & $  0.02260\pm 0.00016\, ( 0.02260^{+0.00031}_{-0.00032} ) $ & $  0.02266\pm 0.00012\, ( 0.02266^{+0.00024}_{-0.00024} ) $ & $  0.02252\pm 0.00011\, ( 0.02252^{+0.00021}_{-0.00022} ) $ \\ 
$ \Omega_\mathrm{c} h^2  $ & $  0.1188\pm 0.0012\, ( 0.1188^{+0.0023}_{-0.0024} ) $ & $  0.1184\pm 0.0011\, ( 0.1184^{+0.0022}_{-0.0022} ) $ & $  0.11874\pm 0.00085\, ( 0.1187^{+0.0016}_{-0.0017} ) $ \\ 
$ \tau_\mathrm{reio}  $ & $  0.072\pm 0.012\, ( 0.072^{+0.023}_{-0.022} ) $ & $  0.070\pm 0.011\, ( 0.070^{+0.022}_{-0.021} ) $ & $  0.0569\pm 0.0071\, ( 0.057^{+0.014}_{-0.014} ) $ \\ 
$ H_0  $ & $  67.67\pm 0.60\, ( 67.7^{+1.2}_{-1.2} ) $ & $  67.69\pm 0.61\, ( 67.7^{+1.2}_{-1.2} ) $ & $  67.64\pm 0.60\, ( 67.6^{+1.2}_{-1.2} ) $ \\ 
$ n_\mathrm{s}  $ & $  0.9763\pm 0.0068\, ( 0.976^{+0.013}_{-0.013} ) $ & $  0.9721\pm 0.0042\, ( 0.9721^{+0.0083}_{-0.0082} ) $ & $  0.9726\pm 0.0032\, ( 0.9726^{+0.0062}_{-0.0066} ) $ \\ 
$ \log(10^{10} A_\mathrm{s})  $ & $  3.072\pm 0.021\, ( 3.072^{+0.041}_{-0.041} ) $ & $  3.074\pm 0.020\, ( 3.074^{+0.040}_{-0.039} ) $ & $  3.048\pm 0.013\, ( 3.048^{+0.025}_{-0.025} ) $ \\

\hline \hline
\end{tabular} }
\end{center}
\caption{Results at 68\% (95\%) CL on cosmological parameters obtained considering different combinations of ACT, Planck-650 and WMAP data in combination with DESI BAO and Pantheon-plus SN distance moduli measurements.}
\label{Tab:Results_ACT_PP}
\end{table*}

\begin{table*}[h]
\begin{center}
\renewcommand{\arraystretch}{1.5}
\resizebox{\textwidth}{!}{
\begin{tabular}{l c c c c c c c c c c c c c c c }
\hline
\textbf{Parameter} & \textbf{ ACT+DESI+DESy5 } & \textbf{ ACT+WMAP+DESI+DESy5 } & \textbf{ ACT+Planck-650+DESI+DESy5 } \\ 
\hline\hline

$ w_0  $ & $  -0.764\pm 0.058\, ( -0.76^{+0.12}_{-0.11} ) $ & $  -0.774\pm 0.057\, ( -0.77^{+0.11}_{-0.11} ) $ & $  -0.759\pm 0.057\, ( -0.76^{+0.11}_{-0.11} ) $ \\ 
$ w_{a}  $ & $  -0.78^{+0.26}_{-0.23}\, ( -0.78^{+0.46}_{-0.50} ) $ & $  -0.71\pm 0.23\, ( -0.71^{+0.44}_{-0.47} ) $ & $  -0.79^{+0.23}_{-0.20}\, ( -0.79^{+0.40}_{-0.45} ) $ \\ 
$ \Omega_\mathrm{b} h^2  $ & $  0.02259\pm 0.00016\, ( 0.02259^{+0.00032}_{-0.00032} ) $ & $  0.02265\pm 0.00012\, ( 0.02265^{+0.00024}_{-0.00023} ) $ & $  0.02251\pm 0.00011\, ( 0.02251^{+0.00021}_{-0.00021} ) $ \\ 
$ \Omega_\mathrm{c} h^2  $ & $  0.1192\pm 0.0012\, ( 0.1192^{+0.0022}_{-0.0024} ) $ & $  0.1188\pm 0.0011\, ( 0.1188^{+0.0022}_{-0.0022} ) $ & $  0.11895\pm 0.00087\, ( 0.1189^{+0.0017}_{-0.0018} ) $ \\ 
$ \tau_\mathrm{reio}  $ & $  0.068\pm 0.011\, ( 0.068^{+0.023}_{-0.022} ) $ & $  0.067\pm 0.011\, ( 0.067^{+0.022}_{-0.022} ) $ & $  0.0554\pm 0.0072\, ( 0.055^{+0.014}_{-0.014} ) $ \\ 
$ H_0  $ & $  66.92\pm 0.55\, ( 66.9^{+1.1}_{-1.1} ) $ & $  66.94\pm 0.55\, ( 66.9^{+1.1}_{-1.1} ) $ & $  66.85\pm 0.57\, ( 66.9^{+1.1}_{-1.1} ) $ \\ 
$ n_\mathrm{s}  $ & $  0.9759\pm 0.0068\, ( 0.976^{+0.014}_{-0.013} ) $ & $  0.9713\pm 0.0040\, ( 0.9713^{+0.0080}_{-0.0077} ) $ & $  0.9720\pm 0.0033\, ( 0.9720^{+0.0064}_{-0.0064} ) $ \\ 
$ \log(10^{10} A_\mathrm{s})  $ & $  3.066\pm 0.021\, ( 3.066^{+0.040}_{-0.040} ) $ & $  3.070\pm 0.021\, ( 3.070^{+0.040}_{-0.041} ) $ & $  3.045\pm 0.013\, ( 3.045^{+0.026}_{-0.025} ) $ \\

\hline \hline
\end{tabular} }
\end{center}
\caption{Results at 68\% (95\%) CL on cosmological parameters obtained considering different combinations of ACT, Planck-650 and WMAP data together with DESI BAO and DESy5 SN distance moduli measurements.}
\label{Tab:Results_ACT_DESy5}
\end{table*}

\begin{table*}[h]
\begin{center}
\renewcommand{\arraystretch}{1.5}
\resizebox{\textwidth}{!}{
\begin{tabular}{l c c c c c c c c c c c c c c c }
\hline
\textbf{Parameter} & \textbf{ SPT+DESI+PP } & \textbf{ SPT+WMAP+DESI+PP } & \textbf{ SPT+Planck-650+DESI+PP } \\ 
\hline\hline

$ w  $ & $  -0.888\pm 0.057\, ( -0.89^{+0.11}_{-0.11} ) $ & $  -0.882\pm 0.056\, ( -0.88^{+0.11}_{-0.11} ) $ & $  -0.860\pm 0.056\, ( -0.86^{+0.11}_{-0.11} ) $ \\ 
$ w_{a}  $ & $  -0.27^{+0.26}_{-0.23}\, ( -0.27^{+0.46}_{-0.50} ) $ & $  -0.29^{+0.25}_{-0.22}\, ( -0.29^{+0.43}_{-0.48} ) $ & $  -0.46^{+0.23}_{-0.21}\, ( -0.46^{+0.42}_{-0.45} ) $ \\ 
$ \Omega_\mathrm{b} h^2  $ & $  0.02226\pm 0.00032\, ( 0.02226^{+0.00063}_{-0.00062} ) $ & $  0.02241\pm 0.00020\, ( 0.02241^{+0.00040}_{-0.00039} ) $ & $  0.02232\pm 0.00016\, ( 0.02232^{+0.00031}_{-0.00030} ) $ \\ 
$ \Omega_\mathrm{c} h^2  $ & $  0.1161^{+0.0018}_{-0.0016}\, ( 0.1161^{+0.0033}_{-0.0036} ) $ & $  0.1163^{+0.0017}_{-0.0015}\, ( 0.1163^{+0.0030}_{-0.0033} ) $ & $  0.1178\pm 0.0012\, ( 0.1178^{+0.0023}_{-0.0024} ) $ \\ 
$ \tau_\mathrm{reio}  $ & $  0.059\pm 0.014\, ( 0.059^{+0.027}_{-0.027} ) $ & $  0.058\pm 0.014\, ( 0.058^{+0.027}_{-0.027} ) $ & $  0.0481^{+0.0085}_{-0.0076}\, ( 0.048^{+0.017}_{-0.016} ) $ \\ 
$ H_0  $ & $  67.25\pm 0.63\, ( 67.3^{+1.2}_{-1.2} ) $ & $  67.32\pm 0.61\, ( 67.3^{+1.2}_{-1.2} ) $ & $  67.43\pm 0.60\, ( 67.4^{+1.2}_{-1.2} ) $ \\ 
$ n_\mathrm{s}  $ & $  0.973\pm 0.015\, ( 0.973^{+0.030}_{-0.029} ) $ & $  0.9690\pm 0.0056\, ( 0.969^{+0.011}_{-0.011} ) $ & $  0.9691\pm 0.0052\, ( 0.969^{+0.010}_{-0.010} ) $ \\ 
$ \log(10^{10} A_\mathrm{s})  $ & $  3.046\pm 0.030\, ( 3.046^{+0.058}_{-0.058} ) $ & $  3.043\pm 0.027\, ( 3.043^{+0.053}_{-0.053} ) $ & $  3.025^{+0.018}_{-0.016}\, ( 3.025^{+0.034}_{-0.034} ) $ \\  

\hline \hline
\end{tabular} }
\end{center}
\caption{Results at 68\% (95\%) CL on cosmological parameters obtained considering different combinations of SPT, Planck-650 and WMAP data together with DESI BAO and Pantheon-plus SN distance moduli measurements.}
\label{Tab:Results_SPT_PP}
\end{table*}

\begin{table*}[h]
\begin{center}
\renewcommand{\arraystretch}{1.5}
\resizebox{\textwidth}{!}{
\begin{tabular}{l c c c c c c c c c c c c c c c }
\hline
\textbf{Parameter} & \textbf{ SPT+DESI+DESy5 } & \textbf{ SPT+WMAP+DESI+DESy5 } & \textbf{ SPT+Planck-650+DESI+DESy5 } \\ 
\hline\hline

$ w_0 $ & $  -0.799\pm 0.061\, ( -0.80^{+0.12}_{-0.12} ) $ & $  -0.797\pm 0.061\, ( -0.80^{+0.12}_{-0.11} ) $ & $  -0.774\pm 0.059\, ( -0.77^{+0.12}_{-0.11} ) $ \\ 
$ w_{a}  $ & $  -0.55\pm 0.27\, ( -0.55^{+0.51}_{-0.54} ) $ & $  -0.56^{+0.28}_{-0.24}\, ( -0.56^{+0.49}_{-0.54} ) $ & $  -0.71^{+0.26}_{-0.22}\, ( -0.71^{+0.46}_{-0.51} ) $ \\ 
$ \Omega_\mathrm{b} h^2  $ & $  0.02225\pm 0.00032\, ( 0.02225^{+0.00062}_{-0.00061} ) $ & $  0.02239\pm 0.00020\, ( 0.02239^{+0.00040}_{-0.00040} ) $ & $  0.02229\pm 0.00015\, ( 0.02229^{+0.00031}_{-0.00030} ) $ \\ 
$ \Omega_\mathrm{c} h^2  $ & $  0.1169^{+0.0017}_{-0.0016}\, ( 0.1169^{+0.0031}_{-0.0034} ) $ & $  0.1169^{+0.0016}_{-0.0014}\, ( 0.1169^{+0.0028}_{-0.0031} ) $ & $  0.1181\pm 0.0012\, ( 0.1181^{+0.0022}_{-0.0024} ) $ \\ 
$ \tau_\mathrm{reio}  $ & $  0.058\pm 0.014\, ( 0.058^{+0.028}_{-0.028} ) $ & $  0.057\pm 0.014\, ( 0.057^{+0.027}_{-0.027} ) $ & $  0.0477^{+0.0086}_{-0.0074}\, ( 0.048^{+0.015}_{-0.018} ) $ \\ 
$ H_0  $ & $  66.56\pm 0.59\, ( 66.6^{+1.2}_{-1.2} ) $ & $  66.63\pm 0.58\, ( 66.6^{+1.1}_{-1.1} ) $ & $  66.66\pm 0.57\, ( 66.7^{+1.1}_{-1.1} ) $ \\ 
$ n_\mathrm{s}  $ & $  0.972\pm 0.015\, ( 0.972^{+0.030}_{-0.029} ) $ & $  0.9681\pm 0.0055\, ( 0.968^{+0.011}_{-0.011} ) $ & $  0.9682\pm 0.0052\, ( 0.968^{+0.010}_{-0.010} ) $ \\ 
$ \log(10^{10} A_\mathrm{s})  $ & $  3.045\pm 0.030\, ( 3.045^{+0.060}_{-0.059} ) $ & $  3.042\pm 0.027\, ( 3.042^{+0.053}_{-0.054} ) $ & $  3.025^{+0.018}_{-0.016}\, ( 3.025^{+0.032}_{-0.036} ) $ \\  

\hline \hline
\end{tabular} }
\end{center}
\caption{Results at 68\% (95\%) CL on cosmological parameters obtained considering different combinations of SPT, Planck-650 and WMAP data together with DESI BAO and DESy5 SN distance moduli measurements.}
\label{Tab:Results_SPT_DESy5}
\end{table*}

\begin{figure*}[htbp!]
    \centering
    \includegraphics[width=\textwidth]{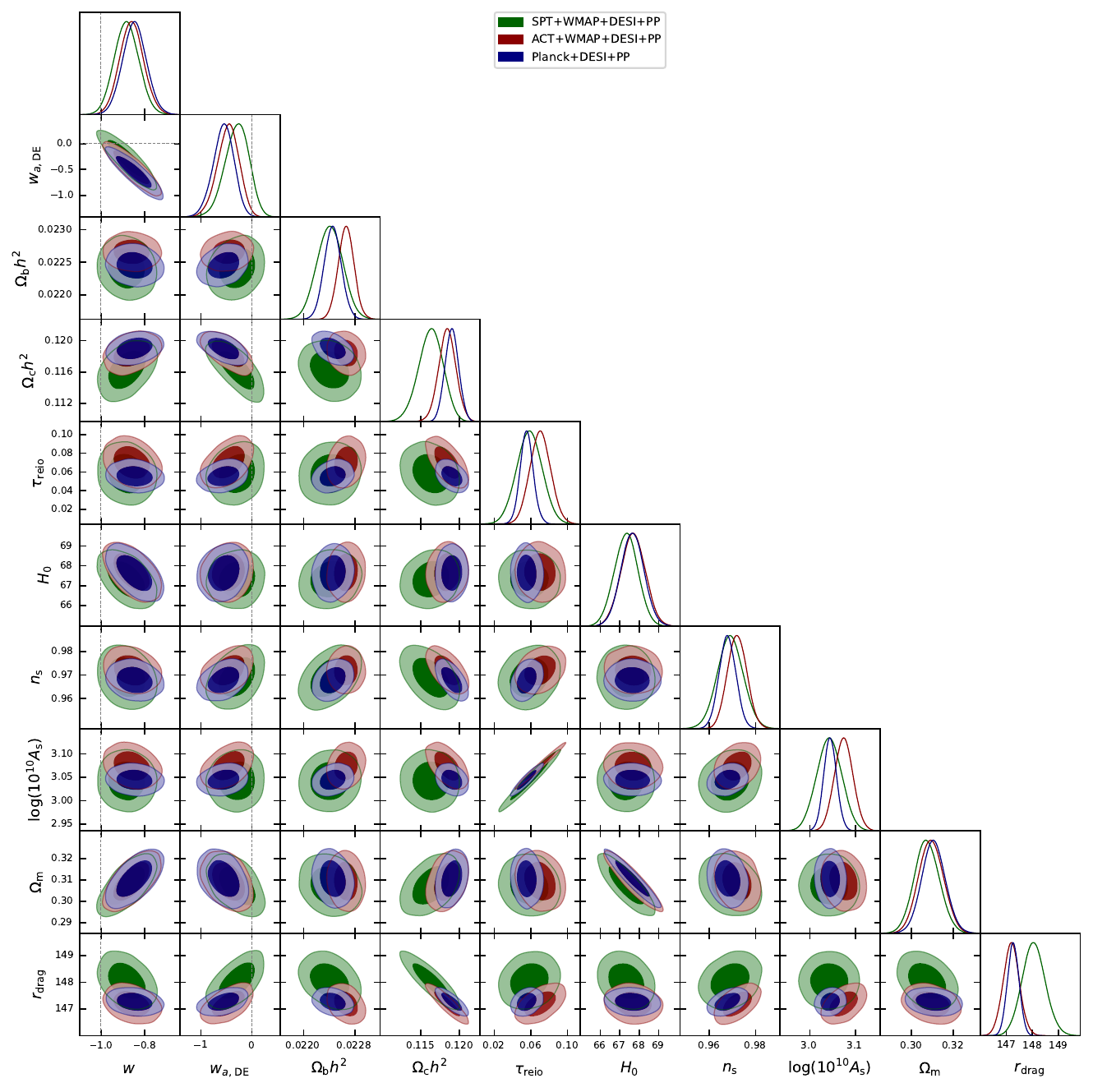}
    \caption{One-dimensional marginalized posterior distributions and two-dimensional joint contours for the most relevant cosmological parameters, inferred from analyzing Planck, ACT, SPT, and WMAP CMB data in combination with DESI BAO and Pantheon-plus SN distance moduli measurements.}
    \label{fig:4}
\end{figure*}

\begin{figure*}[htbp!]
    \centering
    \includegraphics[width=\textwidth]{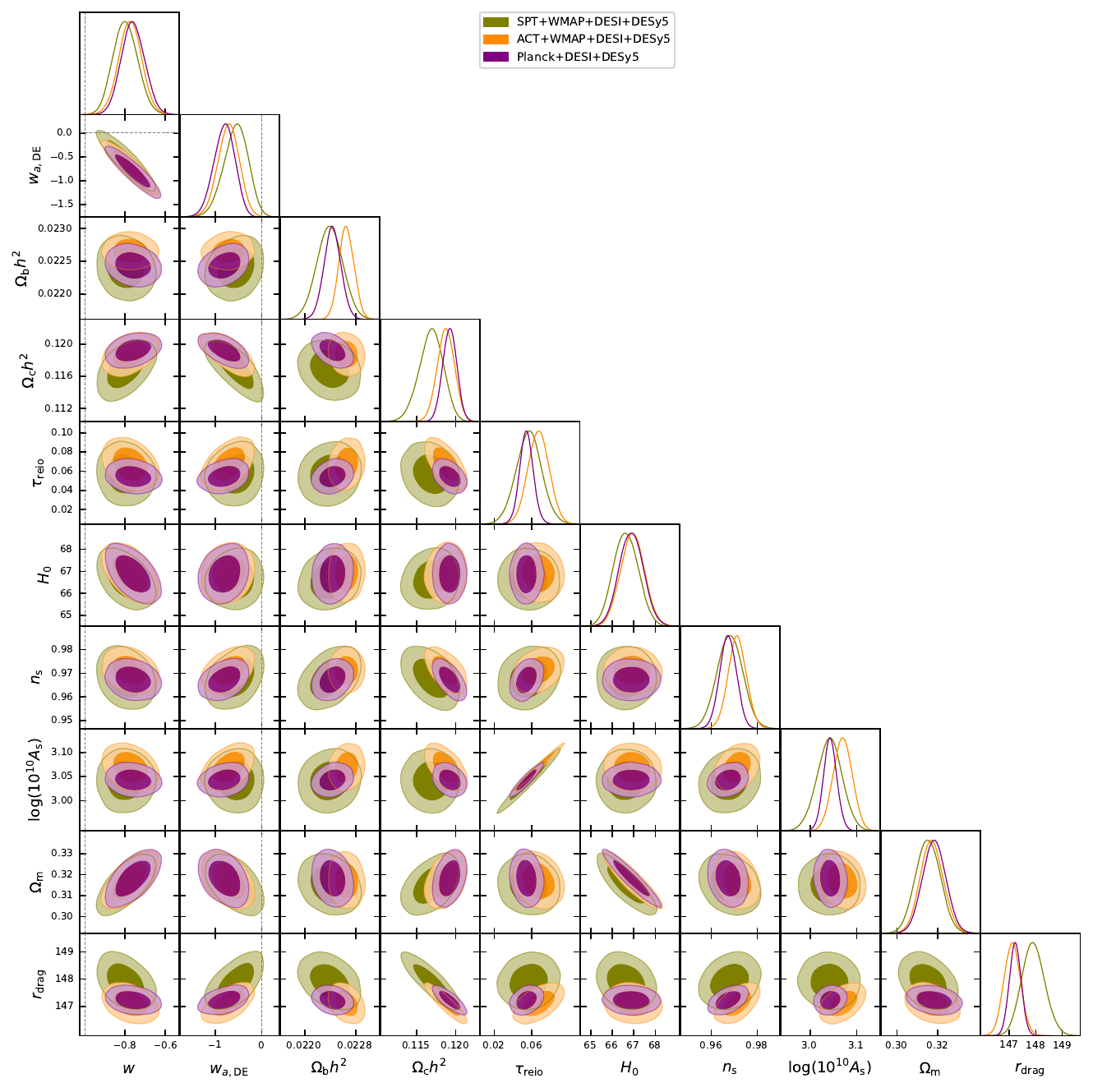}
    \caption{One-dimensional marginalized posterior distributions and two-dimensional joint contours for the most relevant cosmological parameters, inferred from analyzing Planck, ACT, SPT, and WMAP CMB data in combination with DESI BAO and DESy5 SN distance moduli measurements.}
    \label{fig:5}
\end{figure*}

\begin{figure*}[htbp!]
    \centering
    \includegraphics[width=\textwidth]{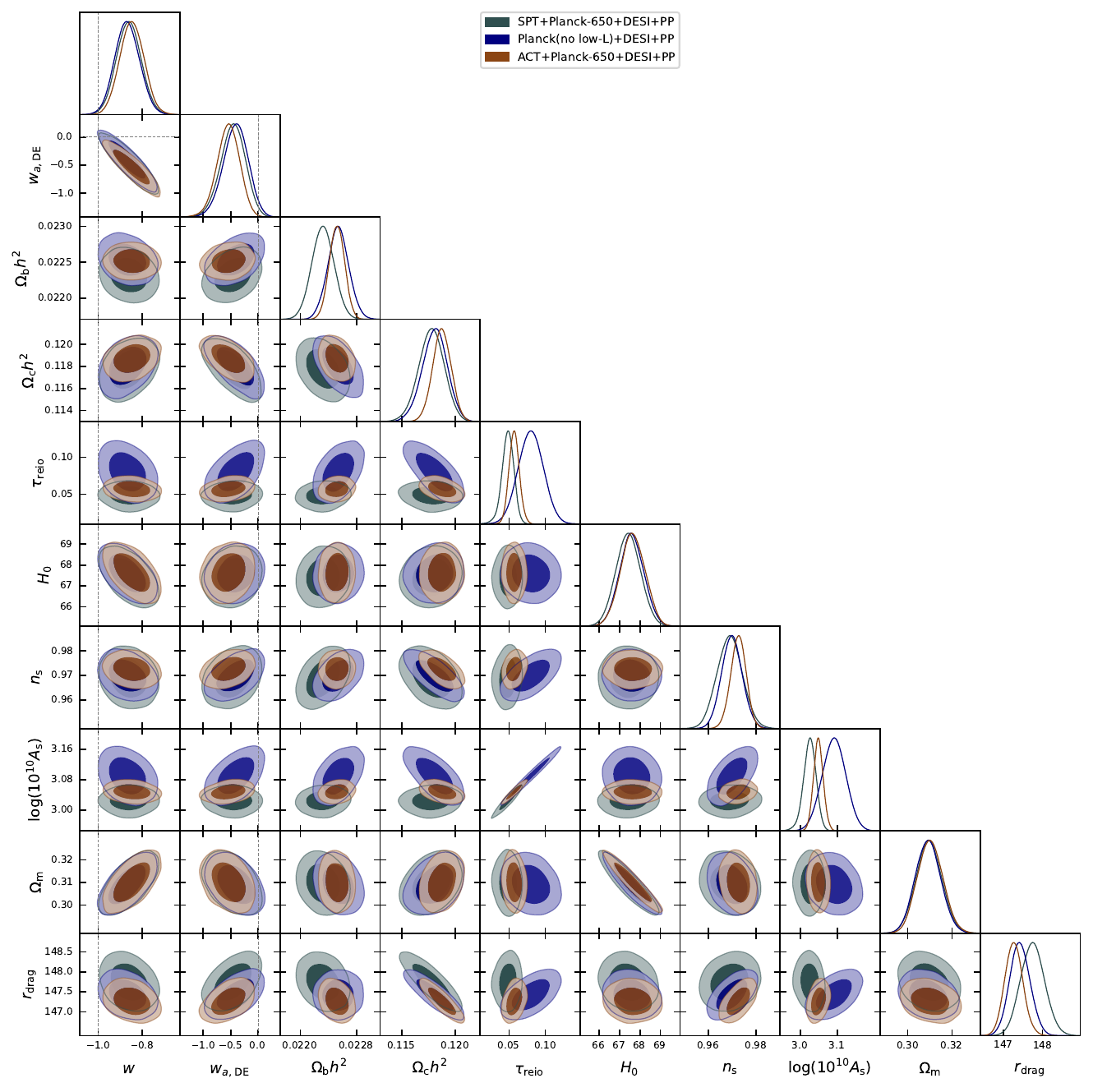}
    \caption{One-dimensional marginalized posterior distributions and two-dimensional joint contours for the most relevant cosmological parameters, inferred from analyzing Planck(no low-L) as well as different combinations of Planck-650, ACT and SPT CMB data in conjunction with DESI BAO and Pantheon-plus SN distance moduli measurements.}
    \label{fig:6}
\end{figure*}

\begin{figure*}[htbp!]
    \centering
    \includegraphics[width=\textwidth]{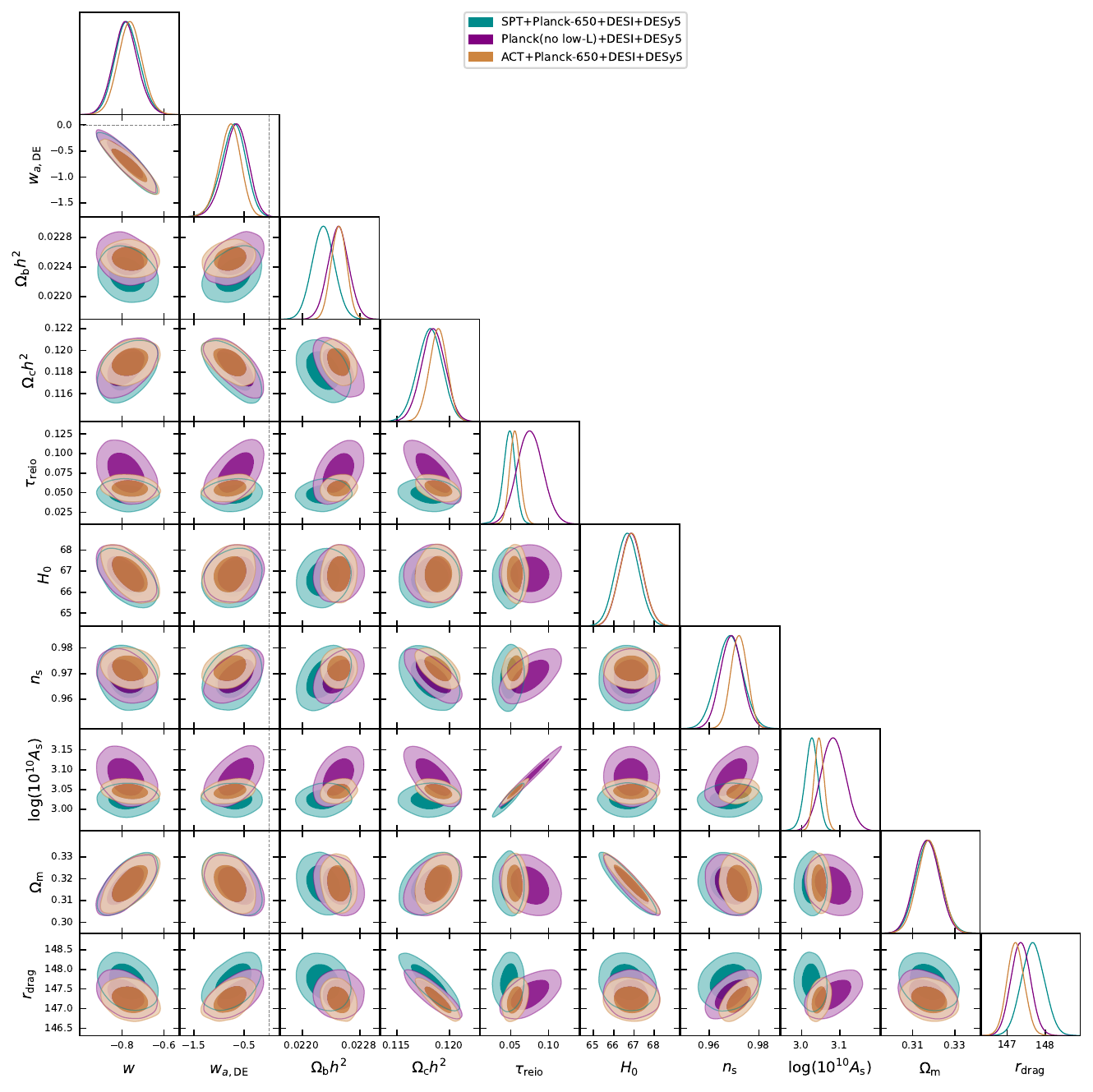}
    \caption{One-dimensional marginalized posterior distributions and two-dimensional joint contours for the most relevant cosmological parameters, inferred from analyzing Planck(no low-L) as well as different combinations of Planck-650, ACT and SPT CMB data in conjunction with DESI BAO and DESy5 SN distance moduli measurements.}
    \label{fig:7}
\end{figure*}

\clearpage

\section{The role of the reionization optical depth and matter density parameters}
\label{appendix:B}

As I showed in the main text of this paper, removing Planck large-scale temperature and polarization data -- and above all E-mode polarization measurements -- or considering CMB experiments different from Planck significantly weakens the preference for dynamical dark energy. As I pointed out in the main manuscript, this is due to at least two different effects. The first is related to the ISW contribution, which mainly affects large-scale temperature modes. Removing the TT data slightly weakens the DDE signal. However, I concluded that the most significant effect comes from removing large-scale E-mode polarization measurements.

As we originally noted in Ref.~\cite{Giare:2023ejv}, Planck low-$\ell$ E-mode polarization data typically favor lower values of the reionization optical depth $\tau$, which correlates with several parameters that are often primarily involved in anomalous cosmological results. For example, after the second data release of DESI BAO measurements, a few independent groups pointed out that such low values of $\tau$ might play a primary role in driving the total neutrino mass towards exceedingly small values~\cite{Sailer:2025lxj,Jhaveri:2025neg}.

Here I argue that large-scale constraints on $\tau$ trigger a domino effect of parameter correlations that plays a key role in strengthening  the evidence for DDE. In particular, due to the well-known degeneracy between $\tau$ and the scalar amplitude $A_s$ -- CMB anisotropies are primarily sensitive to the combination $A_s e^{-2\tau}$ -- a lower $\tau$ leads to a lower inferred $A_s$. This, in turn, impacts the lensing-induced smoothing of the CMB peaks, which depends on the interplay among $(A_s, n_s, \Omega_m)$. To match the observed lensing amplitude, a smaller $A_s$ is typically compensated by a larger $n_s$ and a smaller $\Omega_m$. Since in several dataset combinations the evidence for dynamical dark energy becomes stronger at higher values of $\Omega_m$, this sequence of adjustments can indirectly weaken the DDE signal.

\begin{figure*}[htbp!]
    \centering
    \includegraphics[width=0.7\textwidth]{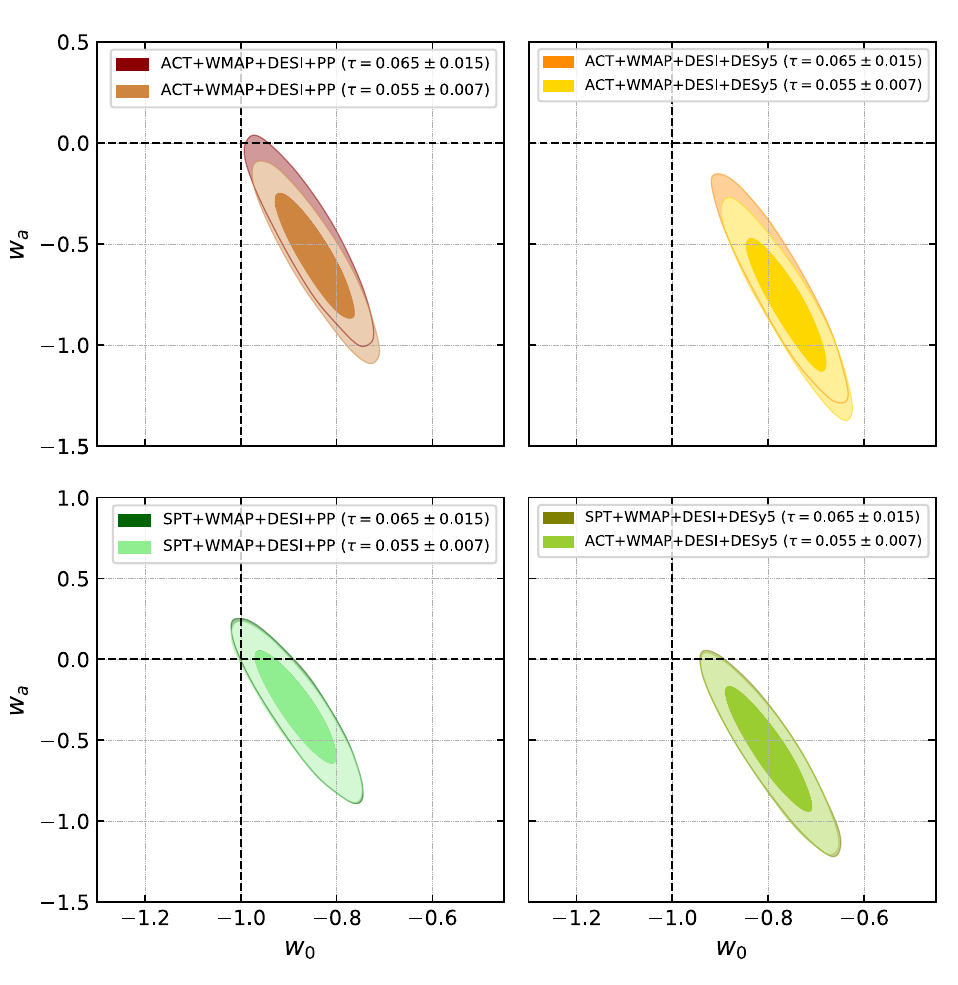}
    \caption{Two-dimensional marginalized contours at 68\% and 95\% CL in the ($w_0$, $w_a$) plane for datasets involving various CMB experiments other than Planck, with different priors on $\tau$ indicated in the legend. The black dashed lines correspond to the standard $\Lambda$CDM values, $w_0 = -1$ and $w_a = 0$.}
    \label{fig:8}
\end{figure*}

To support this argument, I perform two additional analyses:
\begin{enumerate}

    \item When studying constraints from CMB experiments other than Planck, I replace the conservative prior on the reionization optical depth adopted in the main analysis ($\tau = 0.065 \pm 0.015$) with the Planck-driven prior $\tau = 0.055 \pm 0.007$, and check how the results for combinations such as WMAP+ACT and WMAP+SPT change compared to the baseline results discussed in the main text. If the argument is true, a lower value of $\tau$ is expected to strengthen the preference for DDE

    \item When I remove large-scale polarization measurements in combinations that involve Planck CMB data, this automatically leads to larger uncertainties on $\tau$ (with a tendency toward higher values) and $\Omega_m$. To test whether the preference for DDE is weakened by lower values of $\Omega_m$ values, I  impose a prior $\Omega_m = 0.29 \pm 0.01$ and check how the posterior shifts.
\end{enumerate}

The results of the first test are shown in Fig.~\ref{fig:8}. As one can see, in combinations that involve ACT+WMAP data, the preference for DDE is significantly increased when adopting a Planck-driven prior on $\tau$. For example, when considering ACT+WMAP+DESI+PP with a prior $\tau = 0.055 \pm 0.007$, I obtain $w = -0.844 \pm 0.055$ and $w_a = -0.57^{+0.22}_{-0.19}$ (both at 68\% CL). As seen from the figure and by direct comparison with Tab.~\ref{tab:results}, this preference is significantly stronger than in the results presented in the main text. Similarly, replacing PP with DESy5 gives $w = -0.759 \pm 0.056$ and $w_a = -0.80^{+0.23}_{-0.20}$, further strengthening the preference for DDE already present in this combination. That said, the impact of a different $\tau$ prior is much more modest in combinations involving SPT+WMAP. In these cases the results barely change, see also Fig.~\ref{fig:8}. Specifically, imposing a prior $\tau = 0.055 \pm 0.007$ for WMAP+SPT+DESI+PP yields $w = -0.883 \pm 0.056$ and $w_a = -0.30^{+0.24}_{-0.22}$, while for WMAP+SPT+DESI+DESy5 I obtain $w = -0.796 \pm 0.059$ and $w_a = -0.57^{+0.27}_{-0.24}$. The larger uncertainties of SPT mitigate this effect, leading to results similar to those discussed in the main text with the more conservative prior on $\tau$.

\begin{figure*}[htbp!]
    \centering
    \includegraphics[width=0.7\textwidth]{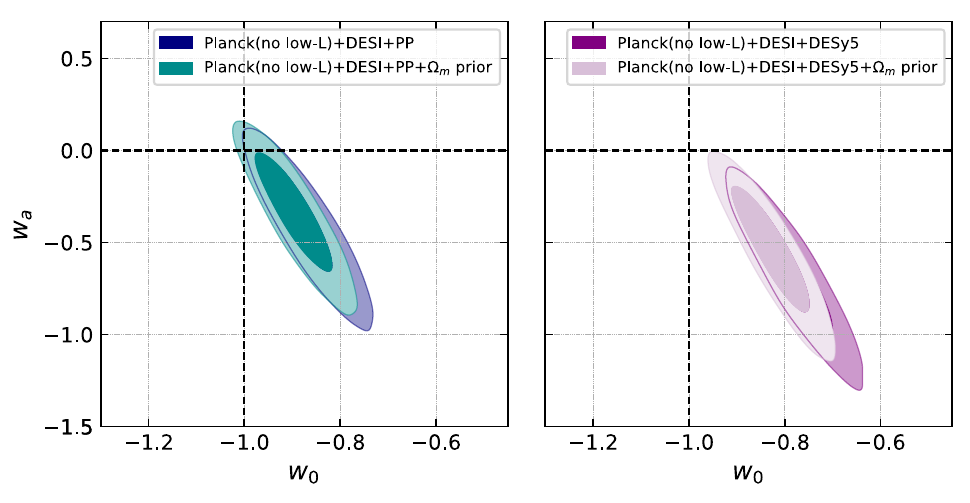}
    \caption{Two-dimensional marginalized contours at 68\% and 95\% CL in the ($w_0$, $w_a$) plane for different combinations of the Planck likelihoods without large-scale temperature and polarization measurements, with and without imposing a prior on $\Omega_m$. The black dashed lines represent the standard $\Lambda$CDM values, $w_0 = -1$ and $w_a = 0$.}
    \label{fig:9}
\end{figure*}

The results of the second test are shown in Fig.~\ref{fig:9}. As seen from the figure, assuming this prior on $\Omega_m$ in combinations of data that involve Planck CMB measurements \textit{without} large-scale temperature and polarization further pushes the contours towards a cosmological constant, decreasing even more the strength of the evidence for DDE. Specifically, imposing a prior $\Omega_m = 0.29 \pm 0.01$ for Planck(no low-L)+DESI+PP gives $w = -0.895 \pm 0.053$ and $w_a = -0.34^{+0.23}_{-0.20}$, while replacing PP with DESy5 for Planck(no low-L)+DESI+DESy5 yields $w = -0.825 \pm 0.054$ and $w_a = -0.55^{+0.25}_{-0.21}$.

These results confirm the key role played by E-mode polarization measurements and the correlations among these quantities, further clarifying the nature of the differences between different subsets of CMB data and the importance of CMB in the analysis.

\end{document}